\documentclass[nohyper,11pt,letterpaper]{JHEP3}
\usepackage[dvips]{epsfig}
\usepackage{epsf,amsfonts,amssymb}
\newcommand{\eqn}[1]{(\ref{#1})}
\newcommand{\be}{\begin{equation}}
\newcommand{\ee}{\end{equation}}
\newcommand{\ben}{\begin{displaymath}}
\newcommand{\een}{\end{displaymath}}
\newcommand{\bea}{\begin{eqnarray}}
\newcommand{\eea}{\end{eqnarray}}
\newcommand{\bean}{\begin{eqnarray*}}
\newcommand{\eean}{\end{eqnarray*}}

\newcommand{\ba}{\begin{array}}
\newcommand{\ea}{\end{array}}
\newcommand{\bi}{\begin{itemize}}
\newcommand{\ei}{\end{itemize}}

\def\l {\lambda}
\def\a {\alpha}

\def\d {\delta}
\def\s {\sigma}

\def\vp {\varphi}

\renewcommand{\O}{\Omega}

\renewcommand{\t}{\theta}

\newcommand{\cale}{\mbox{${\cal E}$}}

\newcommand{\calh}{\mbox{${\cal H}$}}

\newcommand{\call}{\mbox{${\cal L}$}}

\newcommand{\caln}{\mbox{${\cal N}$}}
\newcommand{\calo}{\mbox{${\cal O}$}}
\newcommand{\calp}{\mbox{${\cal P}$}}

\newcommand{\calr}{\mbox{${\cal R}$}}

\newcommand{\bbr}[1]{\mbox{${\mathbb R}^{#1}$}}

\newcommand{\bbi}[1]{\mbox{${\mathbb I}_{#1}$}}

\newcommand{\mkk}{M_\mt{KK}}

\newcommand{\ads}[1]{\mbox{${AdS}_{#1}$}}
\newcommand{\adss}[2]{\mbox{$AdS_{#1}\times {S}^{#2}$}}

\newcommand{\pa}{\partial}
\newcommand{\fc}{\frac}

\newcommand{\sac}{\ , \qquad}
\newcommand{\eg}{{\it e.g.}}
\newcommand{\ie}{{\it i.e.}}

\newcommand{\ra}{\rightarrow}

\newcommand{\beq}{\begin{equation}}
\newcommand{\eeq}{\end{equation}}
\newcommand{\beqr}{\begin{displaymath}}
\newcommand{\eeqr}{\end{displaymath}}
\newcommand{\beqa}{\begin{eqnarray}}
\newcommand{\eeqa}{\end{eqnarray}}
\newcommand{\beqar}{\begin{eqnarray*}}
\newcommand{\eeqar}{\end{eqnarray*}}
\renewcommand{\r}{\rho}

\newcommand{\cN}{{\cal N}}
\newcommand{\cL}{{\cal L}}

\newcommand{\reef}[1]{(\ref{#1})}
\newcommand{\non}{\nonumber}
\newcommand{\pf}{\partial}
\newcommand{\df}{\textrm{d}}
\newcommand{\mt}[1]{\textrm{\tiny #1}}
\newcommand{\mq}{\ensuremath{m_\mt{q}}}      % Quark mass.

\newcommand{\mpi}{\ensuremath{M_{\pi}}}
\newcommand{\Lqcd}{\ensuremath{\Lambda_{\mt{QCD}}}}

\newcommand{\cc}{\langle \bar{\psi} \psi \rangle}
\newcommand{\fpi}{f_\pi}

\def\lt {\lambda}
\def\rt {r}
\def\rhot {\rho}
\def \rvac{r_\mt{vac}}
\def\nc {N_\mt{c}}
\def\nf {N_\mt{f}}
\def\ua {U(1)_\mt{A}}
\def\t6 {T_\mt{D6}}
\def\ut {U_\mt{KK}}
\def\uh {U_\mt{T}}
\def\gym {g_\mt{YM}}
\newcommand{\nonsol}{D4-soliton}
\newcommand{\te}{t_\mt{E}}
\newcommand{\ct}{T_\mt{deconf}}
\newcommand{\tcc}{T_\mt{fund}}
\newcommand{\tb}{\bar{M}}
\newcommand{\msusy}{M_\mt{susy}}
\newcommand{\lm}{\l_\mt{match}}

\title{\LARGE Towards a holographic dual of large-$N_\mathrm{\large c}$ QCD}

\author{Mart\'\i n Kruczenski,$^{a}$
  David Mateos,$^{b}$ Robert C. Myers$\,^{b,c}$ and
  David J. Winters$\,^{b,d}$ \\
  $^a$ Department of Physics, Brandeis University \\
       Waltham, MA 02454, USA \\
  $^b$ Perimeter Institute for Theoretical Physics \\
       Waterloo, Ontario N2J 2W9, Canada \\
  $^c$ Department of Physics, University of Waterloo  \\
       Waterloo, Ontario N2L 3G1, Canada \\
  $^d$ Department of Physics, McGill University \\
       Montr\'eal, Qu\'ebec H3A 2T8, Canada

E-mail: \email{martink@brandeis.edu,
  dmateos@perimeterinstitute.ca, rmyers@perimeterinstitute.ca,
  winters@physics.mcgill.ca}}

\abstract{We study $\nf$ D6-brane probes in the supergravity
  background dual to $\nc$ D4-branes compactified on
  a circle with supersymmetry-breaking boundary conditions.
  In the limit in which the resulting Kaluza--Klein modes decouple,
  the gauge theory reduces to non-supersymmetric,
  four-dimensional QCD with $\nc$ colours and $\nf \ll \nc$ flavours.
  As expected, this decoupling is not fully realised within the
  supergravity/Born--Infeld approximation.
  For \mbox{$\nf=1$} and massless quarks, $\mq =0$,
  we exhibit spontaneous chiral symmetry breaking by a quark
  condensate, $\cc \neq 0$, and find the associated massless `pion' in the
  spectrum. The latter becomes massive for $\mq >0$, obeying the
  Gell-Mann--Oakes--Renner relation: $\mpi^2 = - \mq \cc/ \fpi^2$.
  In the case $\nf >1$ we provide a holographic version of the
  Vafa--Witten theorem, which states that the $U(\nf)$ flavour symmetry
  cannot be spontaneously broken. Further, we find $\nf^2-1$ unexpectedly light
  pseudo-scalar mesons in the spectrum. We argue that these are not
  (pseudo-)Goldstone bosons and speculate on the string mechanism
  responsible for their lightness. We then study the theory at finite
  temperature and exhibit a phase transition associated with a
  discontinuity in $\cc (T)$. D6/$\overline{\mbox{D6}}$ pairs are also
  briefly discussed.}


\preprint{BRX TH-258 \\ 
\tt{hep-th/0311270}}

\begin{document}

%%%%%%%%%%%%%%%%%%%%%%%%%%%%%%%%%%%%%%%%%%%%%%%%%%%%%%%%%%%%%%%%%%%%%%%%%%%%
%%%%%%%%%%%%%%%%%%%%%%%%%%%%%%%%%%%%%%%%%%%%%%%%%%%%%%%%%%%%%%%%%%%%%%%%%%%%
%%%%%%%%%%%%%%%%%%%%%% SECTION  %%%%%%%%%%%%%%%%%%%%%%%%%%%%%%%%%%%%%%%%%%%%
%%%%%%%%%%%%%%%%%%%%%%%%%%%%%%%%%%%%%%%%%%%%%%%%%%%%%%%%%%%%%%%%%%%%%%%%%%%%
%%%%%%%%%%%%%%%%%%%%%%%%%%%%%%%%%%%%%%%%%%%%%%%%%%%%%%%%%%%%%%%%%%%%%%%%%%%%
\section{Introduction and summary of results} \label{intro}

The AdS$_5$/CFT$_4$ correspondence relates string theory on the
near-horizon background of $\nc$ D3-branes to the conformal field
theory living on their worldvolume \cite{Maldacena97}. Inspired by
this correspondence, Witten proposed a construction of the
holographic dual of four-dimensional, pure $SU(\nc)$ Yang--Mills
(YM) theory \cite{Witten98b}. One starts with $\nc$ D4-branes
compactified on a circle of radius $M_\mt{KK}^{-1}$, and further
imposes anti-periodic boundary conditions for the worldvolume
fermions on this circle. Before compactification, the D4-brane
theory is a five-dimensional, supersymmetric $SU(\nc)$ gauge
theory whose field content includes fermions and scalars in the 
adjoint representation of $SU(\nc)$, in addition to the gauge
fields. At energies much below the compactification scale,
$M_\mt{KK}$, the theory is effectively four-dimensional. The
anti-periodic boundary conditions break all of the supersymmetries
and give a tree-level mass to the fermions, while the scalars also
acquire a mass through one loop-effects. Thus at sufficiently low
energies the dynamics is that of four-dimensional, massless
gluons.

The D4-brane system above has a dual description in terms of
string theory in the near-horizon region of the associated (non-supersymmetric) 
supergravity background.
Unfortunately, as observed in \cite{Witten98b}, the Kaluza--Klein
(KK) modes on the D4-brane do not decouple within the supergravity
approximation. For example, the mass of the lightest glueball is
of the same order as the strong coupling scale. In this sense,
there is no energy region, $\Lambda_\mt{QCD} \ll E \ll M_\mt{KK}$,
in which a description in terms of weakly-coupled gluons is
appropriate. Nevertheless, the qualitative features of the
glueball spectrum agree with lattice calculations \cite{COOT98},
and the supergravity description suffices, for example, to exhibit
an area law for Wilson loops \cite{Witten98b}.

Since most of our phenomenological understanding of low-energy QCD
comes from the study of mesons and baryons, it is interesting to
construct a holographic dual of a gauge theory whose low-energy
degrees of freedom consist not only of gluons but also of
fundamental quarks. In the context of AdS/CFT, fundamental matter
can be added to the gauge theory by introducing D-brane probes in
the dual supergravity background
\cite{AFM98,KR01,BDFLM01,BDFM01,KK02, KKW02, WH03, Ouyang03}. This has
been recently exploited to study mesons holographically in several
examples of gauge/gravity duals \cite{KMMW03, SS03, BEEGK03, NPR03}. 
The goal of this paper is to study a gauge/gravity dual in which the
gauge theory reduces to non-supersymmetric, four-dimensional QCD
in the limit in which the KK modes decouple. The construction is
as follows.

Consider the D4/D6 system with the branes oriented as described
by the following array:
\be
\begin{array}{rccccccccccl}
\nc \,\, \mbox{D4:}\,\,\, & 0 & 1 & 2 & 3 & 4 & \_ & \_ & \_ & \_ & \_ & \, \\
\nf \,\, \mbox{D6:}\,\,\, & 0 & 1 & 2 & 3 & \_ & 5 & 6 & 7 & \_ & \_ & \, .
\ea
\label{intersection}
\ee
Note that the D4- and the D6-branes may be separated from each other
along the 89-directions. This system is T-dual to the D3/D5
intersection, and in the decoupling limit for the D4-branes
\cite{IMSY98} provides a non-conformal version of the AdS/dCFT
correspondence \cite{KR01, DFO01}. On the gauge theory side one has
a supersymmetric, five-dimensional $SU(\nc)$ gauge theory coupled to a
four-dimensional defect. The entire system is invariant under
eight supercharges, that is, $\caln =2$ supersymmetry in
four-dimensional language. The degrees of freedom localized on
the defect are $\nf$ hypermultiplets in the fundamental representation
of $SU(\nc)$, which arise from the open strings connecting the D4- and
the D6-branes. Each hypermultiplet consists of two Weyl fermions
of opposite chiralities, $\psi_\mt{L}$ and $\psi_\mt{R}$,
and two complex scalars.

As discussed above, identifying the 4-direction with period
$2\pi/M_\mt{KK}$, and with anti-periodic boundary conditions for
the D4-brane fermions, breaks all of the supersymmetries and
renders the theory effectively four-dimensional at energies $E \ll
M_\mt{KK}$. Further, the adjoint fermions and scalars become
massive. Now, the bare mass of each hypermultiplet, $\mq$, is
proportional to the distance between the corresponding D6-brane
and the D4-branes. Even if these bare masses are zero, we expect
loop effects to induce a mass for the scalars in the fundamental
representation. Generation of a mass for the fundamental fermions
is, however, forbidden by a chiral, $\ua$ symmetry that rotates
$\psi_\mt{L}$ and $\psi_\mt{R}$ with opposite
phases.\footnote{This symmetry is broken by instanton
effects, but these vanish in the large-$\nc$ limit in which
we work.} Therefore, at low energies, we expect to
be left with a four-dimensional $SU(\nc)$ gauge theory coupled to
$\nf$ flavours of fundamental quark.

In the dual string theory description, the D4-branes are again
replaced by their supergravity background. In the so-called `probe
limit', $\nf \ll \nc$, the backreaction of the D6-branes on this
background is negligible and hence they can be treated as probes.
The D6-brane worldvolume fields are dual to
gauge-invariant field theory operators constructed with at least two
hypermultiplet fields, that is, meson-like operators; of
particular importance here will be the quark bilinear operator,
$\bar{\psi} \psi \equiv \bar\psi_i \psi^i$, where 
$\psi^i = \psi_\mt{L}^i + \psi_\mt{R}^i$ and $i=1, \ldots , \nf$ 
is the flavour index. In
the string description the $\ua$ symmetry is nothing but the
rotation symmetry in the 89-plane. In addition to acting on the
fermions as explained above, this symmetry also acts on the
adjoint scalar $X=X^8 + i X^9$ by a phase rotation.

Having presented the general construction, we now summarise our main
results. 

We begin in section \ref{broken} by considering the case of a single 
D6-brane.  For $\mq=0$, we expect the chiral $\ua$ symmetry to be 
spontaneously broken by a chiral condensate, \mbox{$\cc \neq 0$}, and there 
to be an associated massless, pseudo-scalar Goldstone boson in the 
spectrum, as in QCD with one massless flavour. In QCD, this is the 
$\eta'$ (which becomes massless in the large-$\nc$ limit), but, in an 
abuse of language, we will refer to it as a `pion'. 
We are indeed able to show that the string description provides a
holographic realisation of this physics, by showing that the $\ua$ 
symmetry is spontaneously broken by the brane embedding, as in
\cite{BEEGK03}. Moreover, we numerically determine the chiral condensate 
for an arbitrary quark mass. We find \mbox{$\cc(\mq=0)\neq 0$}, as expected,
and $\cc(\mq) \propto 1/\mq$ for $\mq \ra \infty$, again in agreement 
with field theory expectations (which we briefly review).
 
In sections \ref{fluct} and \ref{pseudo} we study the meson spectrum. 
We first show analytically, in section \ref{fluct}, that at $\mq=0$ 
there is exactly one normalisable, massless pseudo-scalar 
(in the four-dimensional sense) fluctuation of the D6-brane. 
This open-string mode corresponds 
precisely to the Goldstone mode of the D6-brane embedding. If $\mq>0$, 
this pion becomes a pseudo-Goldstone boson. We are able to show, 
analytically, that its squared mass scales linearly with the quark
mass, $\mpi^2 \propto \mq$, in the limits of small (section
\ref{pseudo}) and large (section \ref{fluct}) quark mass. With 
numerical calculations we then confirm that, in fact, such a linear 
relation holds for all quark masses. 
For small $\mq$ this linear relation is in perfect agreement with the 
Gell-Mann--Oakes--Renner (GMOR) relation \cite{GMOR68},
\be 
\mpi^2 = - \fc{\mq \, \cc}{\fpi^2} \,,
\label{GMOR} 
\ee
which gives the first term in the expansion of the pion mass around 
$\mq=0$. In section \ref{pseudo}, we compute the chiral condensate and the 
pion decay constant analytically (at $\mq=0$), and with these results 
we are able to verify that, for small quark mass, the mass of our pion
precisely satisfies the GMOR relation. In the opposite limit, 
$\mq \ra \infty$, we show in section \ref{fluct} that 
\be
\mpi^2 = a \, \fc{\mq \mkk}{\gym^2 \nc} \,,
\label{linear} 
\ee
where $a$ is a pure number; for the lightest mesons, we find $a \sim 20$.

The remaining mesons (D6-brane excitations) are studied in section
\ref{fluct}. For $\mq \lesssim \msusy$, where 
\be
\msusy = \gym^2 \nc \mkk \,,
\label{msusy}
\ee 
they all have masses of order the compactification scale, $\mkk$, and 
so the (pseudo-)Goldstone boson dominates the infrared physics in the
regime of small $\mq$. Of course, this result signals the lack of
decoupling between the KK and the QCD scales within the 
supergravity/Born--Infeld approximation, as expected. For 
$\mq \gtrsim \msusy$, supersymmetry is approximately restored, 
the spectrum exhibits the corresponding degeneracy, and all meson
masses obey a formula like \eqn{linear} with appropriate values of
$a$. The fact that supersymmetry is restored at $\mq \sim \msusy$ 
suggests that the effective mass of some of the microscopic degrees 
of freedom must be at least of order $\msusy$. We close section 
\ref{fluct} with an examination of certain stability issues for the
D6-brane embeddings. 

In section \ref{vafa}, for the case of multiple flavours, $\nf >1$,
we provide a holographic version of the Vafa--Witten theorem
\cite{VW84}, which states that the $U(\nf)$ flavour symmetry
cannot be spontaneously broken if $\mq >0$. In the holographic
description this is realised by the fact that the $\nf$ D6-branes
must be coincident in order to minimise their energy.

In section \ref{fintem}, we examine the theory at finite
temperature. If $\mq> \msusy$, the theory exhibits two
phase transitions: the first one is the well-known
confinement/deconfinement transition at $T = \ct \sim \mkk$
\cite{Witten98b}, and the second one is a phase transition at 
\be
T = \tcc  \sim \sqrt{\fc{\mq \mkk}{\gym^2 \nc}} > \ct \,, 
\ee
characterised by a discontinuity in the chiral condensate, 
$\cc (T)$, and in the specific heat. If $\mq < \msusy$ 
the second transition does not occur as a separate phase transition. 

In section \ref{Dbar} we discuss six-brane embeddings 
that, from the brane-construction viewpoint, correspond not to
D4/D6 intersections but to intersections of D4-branes with
D6/$\overline{\mbox{D6}}$ pairs. From the gauge theory viewpoint these
configurations correspond to having a defect/anti-defect pair. 

We close with a brief discussion of our results in section
\ref{discus}. In particular, one result of the analysis of the
$\nf>1$ case in section \ref{vafa} is the existence of $\nf^2$
massless, pseudo-scalar mesons in the gauge theory spectrum if $\mq=0$.
(We expect the inclusion of sub-leading effects in the $1/\nc$
expansion to generate small masses for these particles.) Although
it is tempting to interpret these particles as Goldstone bosons of
a putative, spontaneously-broken $U(\nf)_\mt{A}$ chiral symmetry,
we argue that this interpretation cannot be correct: even at large
$\nc$, only one of them is a true Goldstone boson. 
Thus we are unable to find any obvious reasons, from the viewpoint of
the four- or five-dimensional gauge theory, for these
scalars to be light. We speculate that, from the string
viewpoint, ten-dimensional $U(\nf)$ gauge invariance (as opposed
to just seven-dimensional gauge invariance on the D6-branes) is
responsible for their lightness.

%%%%%%%%%%%%%%%%%%%%%%%%%%%%%%%%%%%%%%%%%%%%%%%%%%%%%%%%%%%%%%%%%%%%%%%%%%%%%%
%%%%%%%%%%%%%%%%%%%%%%%%%%%%%%%%%%%%%%%%%%%%%%%%%%%%%%%%%%%%%%%%%%%%%%%%%%%%%%
%%%%%%%%%%%%%%%%%%%%%% SECTION  %%%%%%%%%%%%%%%%%%%%%%%%%%%%%%%%%%%%%%%%%%%%%
%%%%%%%%%%%%%%%%%%%%%%%%%%%%%%%%%%%%%%%%%%%%%%%%%%%%%%%%%%%%%%%%%%%%%%%%%%%%%%
%%%%%%%%%%%%%%%%%%%%%%%%%%%%%%%%%%%%%%%%%%%%%%%%%%%%%%%%%%%%%%%%%%%%%%%%%%%%%%
\section{Chiral symmetry breaking from D6-brane embeddings}
\label{broken}

In this section we will show how the spontaneous breaking of the
$\ua$ chiral symmetry expected from the field theory side is
realized in the string description.

%%%%%%%%%%%%%%%%%%%%%%%%%%%%%%%%%%%%%%%%%%%%%%%%%%%%%%%%%%%%%%%%%%%%
\subsection{The D4-soliton background}
%%%%%%%%%%%%%%%%%%%%%%%%%%%%%%%%%%%%%%%%%%%%%%%%%%%%%%%%%%%%%%%%%%%%

The type IIA supergravity background dual to $\nc$ D4-branes
compactified on a circle with anti-periodic boundary conditions
for the fermions takes the form
\begin{eqnarray}
ds^{2} &=& \left(\frac{U}{R}\right)^{3/2} \left( \eta_{\mu \nu} \,
dx^\mu dx^\nu + f(U) d\tau^{2} \right) + \left(
\frac{R}{U}\right)^{3/2} \frac{dU^{2}}{f(U)} +
R^{3/2} U^{1/2} \, d\Omega_{\it 4}^{2} \,, \label{metric} \\
e^{\phi} &=& g_s \left( \frac{U}{R}\right)^{3/4}
\sac F_{\it 4} = \frac{\nc}{V_{\it 4}} \, \varepsilon_{\it 4} \sac
f(U) = 1-\frac{\ut^{3}}{U^{3}} \,.
\label{metric1}
\end{eqnarray}
The coordinates $x^\mu=\{ x^0, \ldots , x^3\}$ parametrize the
four non-compact directions along the D4-branes, as in
\eqn{intersection}, whereas $\tau$ parametrizes the circular
4-direction on which the branes are compactified. $d\Omega_{\it
4}^2$ and $\varepsilon_{\it 4}$ are the $SO(5)$-invariant line
element and volume form on a unit four-sphere, respectively, and
$V_{\it 4}=8\pi^2/3$ is its volume. $U$ has dimensions of length
and may be thought of as a radial coordinate in the
56789-directions transverse to the D4-branes. To avoid a conical
singularity at $U=\ut$, $\tau$ must be identified with period
\be 
\d \tau = \fc{4 \pi}{3} \, \fc{R^{3/2}}{\ut^{1/2}} \,.
\label{deltatau} 
\ee
Following the nomenclature of \cite{soliton}, we refer to this
solution as the \nonsol.

This supergravity solution above is regular everywhere and is
completely specified by the string coupling constant, $g_s$, the
Ramond--Ramond flux quantum (\ie, the number of D4-branes), $\nc$, and 
the constant $\ut$. The remaining
parameter, $R$, is given in terms of these quantities and the string
length, $\ell_s$, by
\be 
R^3 =  \pi g_s \nc\,\ell_s^3\,. \label{R} 
\ee
If $\ut$ is set to zero, the solution (\ref{metric},
\ref{metric1}) reduces to the extremal, 1/2-supersymmetric
D4-brane solution. Hence we may say that $\ut$ characterises the
deviation of the \nonsol\ from extremality.

The $SU(\nc)$ field theory dual to (\ref{metric}, \ref{metric1}) is
defined by the compactification scale, $\mkk$, below which the
theory is effectively four-dimensional, and the four-dimensional
coupling constant {\it at} the compactification scale, $\gym$. These are
related to the string parameters by
\be 
\mkk = \fc{3}{2}\fc{\ut^{1/2}}{R^{3/2}} =
\fc{3}{2 \sqrt{\pi}}\fc{\ut^{1/2}}{{( g_s \nc)}^{1/2} \ell_s^{3/2}} 
\sac \gym^2 =3 \sqrt{\pi} \left( \fc{g_s\ut}{\nc \ell_s} \right)^{1/2} \,.
\label{parameters} 
\ee
The first equation follows from the fact that $\tau$ is 
directly identified with the compact direction in the gauge theory,
so $\mkk \equiv {2\pi/\d\tau}$. The second equation follows from 
the fact that $g_{\it{5}}^2 = (2\pi)^2 g_s \ell_s$ and the relation between 
the four- and five-dimensional coupling 
constants, $\gym^2 = g_{\it{5}}^2 / \d \tau$, which can be seen  
by expanding the Born--Infeld action of the D4-brane and adopting 
the standard normalization for the gauge field kinetic term, 
$-F^2 / 4\,g_{\it{5}}^2$.

It is useful to invert these relations to express the string
parameters in terms of the gauge theory ones:
\be
R^3 = {1\over2} \fc{\gym^2 \nc \, \ell_s^2}{\mkk} \sac
g_s = {1\over2\pi} \fc{\gym^2}{\mkk \ell_s} \sac
\ut = {2\over9} \gym^2 \nc \, \mkk \ell_s^2 \,.
\label{inverse}
\ee
The string length will cancel in any calculation of a physical
quantity in the field theory. For example, the QCD string tension 
is\footnote{This is the tension of a string lying at $U=\ut$
and extending along $x$. It can also be computed by projecting such a
string onto the boundary and integrating $\langle F^2\rangle$ across
a transverse section of the string \cite{adsprop}.}
\be
\sigma =  \frac{1}{2\pi\ell_s^2} \left. \sqrt{-G_{tt} G_{xx}}
\right|_{U=\ut} = \frac{1}{2\pi \ell_s^2}
\left(\frac{\ut}{R}\right)^{3/2} = {2\over27\pi} \gym^2 \nc \, \mkk^2 \,.
\ee
Now we may ask in what situation the \nonsol\  provides a reliable
background in which one can study the dual field theory using
only classical supergravity. First, we must require that the curvature is
everywhere small compared to the fundamental string tension. This
ensures that higher derivative string corrections to the low
energy equations of motion are negligible. The maximum curvature
in the geometry given by \reef{metric} occurs at precisely $U=\ut$,
where the curvatures are of the order $(\ut R^3)^{-1/2}$. Hence we
require
\be
{\ut^{1/2} R^{3/2}\over\ell_s^2}  \simeq \gym^2 \nc  \gg 1\,,
\label{hoof}
\ee
where we have applied the results in eq.~\reef{inverse}. Therefore
the restriction to small curvatures corresponds to a
large 't Hooft coupling in the effective four-dimensional gauge
theory, precisely as in the conventional AdS/CFT correspondence
\cite{Maldacena97}. Further, to suppress string loop effects, we
must also require that the local string coupling, $e^\phi$, be small. Using
eqs.~\reef{metric1} and \eqn{inverse} we see that, for finite values 
of the gauge theory parameters, the inequality $e^\phi \ll 1$ can
only be satisfied up to some critical radius
\be
U_\mt{crit} \simeq \fc{\nc^{1/3} \mkk \ell_s^2}{\gym^2} \,.
\label{crit}
\ee
Beyond this radius, the M-theory circle opens up to
reveal an \adss{7}{4} background (with identifications).
Similarly, at the corresponding high-energy scales, the
five-dimensional Yang--Mills theory reveals a UV completion in
terms of the (2,0) theory compactified on a circle. In the present
context, we naturally demand that $U_\mt{crit}\gg\ut$, which,
using \eqn{crit} and \eqn{inverse}, reduces to
\be 
\gym^4\ll{1\over\gym^2\nc}\ll1\,, 
\label{small} 
\ee
where the second inequality follows from eq.~\eqn{hoof}. 
Eqs.~\eqn{hoof} and \eqn{small} imply that the supergravity analysis 
in the \nonsol\ background is reliable in precisely the
strong-coupling regime of the 't Hooft limit of the four-dimensional 
gauge theory: $\gym \ra 0$, $\nc \ra \infty$, $\gym^2 \nc$ fixed and
large.

%%%%%%%%%%%%%%%%%%%%%%%%%%%%%%%%%%%%%%%%%%%%%%%%%%%%%%%%%%%%%%%%%%%%
\subsection{D6-brane embeddings}
%%%%%%%%%%%%%%%%%%%%%%%%%%%%%%%%%%%%%%%%%%%%%%%%%%%%%%%%%%%%%%%%%%%%

We are now ready to study the embedding of a D6-brane probe in the
\nonsol\ geometry. Asymptotically (as $U\rightarrow \infty$), the
D6-brane is embedded as described by the array \eqn{intersection}.
The analysis is greatly simplified by introducing isotropic
coordinates in the 56789-directions. Towards this end, we first
define a new radial coordinate, $\rho$, related to $U$ by
\beq 
U(\rho) = \left(\rho^{3/2} +
\frac{\ut^3}{4\rho^{3/2}}\right)^{2/3} \,, 
\label{isomer}
\eeq
and then five coordinates $\vec{z}=(z^5, \ldots, z^9)$ such that
$\rho = |\vec{z}|$ and $d\vec{z} \cdot d\vec{z} = d\rho^2 + \rho^2
\, d\Omega_{\it 4}^2$. In terms of these coordinates the metric
\eqn{metric} becomes
\be
ds^{2} = \left(\frac{U}{R}\right)^{3/2}
\left( \eta_{\mu \nu} \, dx^\mu dx^\nu + f(U) d\tau^{2} \right) +
K(\rho) \, d\vec{z} \cdot d\vec{z} \,,
\ee
where
\be K(\rho) \equiv \fc{R^{3/2} U^{1/2}}{\rho^2}\,. 
\ee
Here $U$ is now thought of as a function of $\rho$. Finally, to
exploit the symmetries of the D6-brane embedding we seek, we
introduce spherical coordinates $\l, \Omega_{\it 2}$ for the
$z^{5,6,7}$-space and polar coordinates $r, \phi$ for the
$z^{8,9}$-space. The final form of the D4-brane metric is then
\be
ds^{2} = \left(\frac{U}{R}\right)^{3/2} \left( \eta_{\mu \nu} \,
dx^\mu dx^\nu + f(U) d\tau^{2} \right) + K(\rho) \, \left(
d\lambda^2 + \lambda^2 \, d\Omega_{\it 2}^2 + dr^2 + r^2 \,
d\phi^2 \right) \,, \label{isometric}
\ee
where $\rho^2 = \l^2 + r^2$.

In these coordinates the D6-brane embedding takes a particularly
simple form. We use $x^\mu$, $\l$ and $\Omega_{\it 2}$ (or $\s^a$,
$a=0,\ldots,6$, collectively) as worldvolume coordinates.  The
D6-brane's position in the 89-plane is specified as $r=r(\l)$, $\phi
= \phi_0$, where $\phi_0$ is a constant. Note that $\l$ is the
only variable on which $r$ is allowed to depend, by translational
and rotational symmetry in the 0123- and 567-directions,
respectively. We also set $\tau=\mbox{constant}$ in the following,
which corresponds to a single D6-brane localized in the circle
direction. We will study configurations in which this condition is
relaxed in section \ref{Dbar}.

With this ansatz for the embedding, the induced metric on the
D6-brane, $g_{ab}$, takes the form
\be 
ds^2(g) = \left(\frac{U}{R}\right)^{3/2} \, \eta_{\mu \nu} \,
dx^\mu dx^\nu + K(\rho) \, \left[ \left( 1+\dot{r}^2 \right)
d\lambda^2 + \lambda^2 \, d\Omega_{\it 2}^2 \right] \,,
\label{induced} 
\ee
where $\dot{r}\equiv\partial_\lambda r$. The D6-brane action
becomes
\be
S_{D6} = -\fc{1}{(2\pi)^6 \ell_s^7} 
\int \df^7\s\, e^{-\phi} \sqrt{-\det g} =
-\t6 \int \df^7\s\, \sqrt{h} 
\left( 1 + \fc{\ut^3}{4\rho^3} \right)^2 \,
\l^2 \sqrt{1 + \dot{r}^2} \,,
\label{eq:D6action}
\ee
where $T_\mt{D6}= 2\pi/g_s(2\pi \ell_s)^7$ is the six-brane
tension and $h$ is the determinant of the metric on the round unit
two-sphere. Recall that $\rho$ is a function of $\l$ both
explicitly and through its dependence on $r(\l)$. Hence the
equation of motion for $r(\l)$ is
\beq
\frac{d}{d\lambda} \left[\left(1+\frac{\ut^3}{4\rho^3}\right)^2
  \lambda^2 \frac{\dot{r}}{\sqrt{1+\dot{r}^2}} \right]
= -\frac{3}{2} \frac{\ut^3}{\rho^5}
\left(1+\frac{\ut^3}{4\rho^3}\right)
\lambda^2 \, r \, \sqrt{1+\dot{r}^2} \,.
\label{embedeq}
\eeq
Note that  $r(\l)=r_0$, where $r_0$ is a constant, is a solution in 
the supersymmetric limit ($\ut=0$), as in
\cite{KK02,KMMW03}. This reflects the BPS nature of the system,
which implies that there is no force on the D6-brane regardless
of its position in the 89-plane. In particular, then, the solution
with $r_0 =0$ preserves the $\ua$ rotational symmetry in the
89-directions. If $\ut\neq 0$ the force on the D6-brane no longer
vanishes and causes it to bend as dictated by the equation of
motion above. We will see below that in this case there are no
(physical) solutions that preserve the $U(1)_\mt{A}$ symmetry.

If $\ut\neq 0$, the analysis is facilitated by rescaling to {\it
dimensionless} variables as follows:
\beq
\l \to \ut\l \sac r \to \ut r \sac \rho \to
\ut\rho \,, 
\label{tildes}
\eeq
in terms of which equation \eqn{embedeq} becomes
\beq
\frac{d}{d\lt} \left[\left(1+\frac{1}{4\rhot^3}\right)^2 \lt^2
\frac{\dot{\rt}}{\sqrt{1+\dot{\rt}^2}} \right] = -\frac{3}{2}
\frac{1}{\rhot^5} \left(1+\frac{1}{4\rhot^3}\right) \lt^2 \, \rt
\, \sqrt{1+\dot{\rt}^2} \,. 
\label{resc}
\eeq
We seek solutions such that the asymptotic separation of the
D6-brane and the D4-branes, $L$, is finite. That is, as
$\lt\rightarrow\infty$, $\rt(\lt) \rightarrow \rt_\infty$ with
\be
\rt_\infty = \fc{L}{\ut} \,.
\label{L}
\ee
In the region $\lt\rightarrow\infty$ we have $\dot{\rt}\rightarrow 0$
and $\rhot \simeq \lt$. Under these conditions we can linearize equation
\eqn{resc} with the result
\beq
\frac{d}{d\lt} \left[ \left. \lt \right.^2 \dot{\rt} \right]
\simeq -\frac{3}{2} \frac{1}{\left.\lt \right.^3}  \rt \,,
\eeq
the solution to which is
\beq
\rt(\lt) = A \, \frac{1}{\sqrt{\lt}} J_{-1/3} \left(
\frac{2^{1/2}3^{-1/2}}{\left. \lt \right.^{3/2}} \right) + B \,
\frac{1}{\sqrt{\lt}} J_{1/3} \left( \frac{2^{1/2}3^{-1/2}}{\left.
\lt \right.^{3/2}} \right) \,,
\eeq
where $J_{\nu}$ are Bessel functions and $A$ and $B$ are arbitrary
constants. For large $\lt$, the function multiplied by $A$ tends
to a constant, whereas that multiplied by $B$ behaves as
$1/\lt$. This means that for large $\l$ we have
\be
\rt(\l) \simeq \rt_\infty + \fc{c}{\lt} \,,
\label{rt}
\ee
where the constants $\rt_\infty$ and $c$ are related to the quark mass
and the chiral condensate, as we now show.

The bare quark mass is given by the string tension times the
{\it asymptotic} distance between the D4- and the 
D6-brane.\footnote{This mass is trivially derived for the 
brane array \reef{intersection} in asymptotically flat space. 
With supersymmetry ($\ut =0$), this bare mass persists in the 
decoupling limit and is then inherited by the compactified theory 
with supersymmetry-breaking boundary conditions, since setting 
$\ut \neq 0$ does not alter the asymptotic properties of the model
that, as usual in AdS/CFT-like dualities, determine the gauge theory
parameters.} 
Taking \eqn{L} into account, we have
\beq
\mq = \frac{L}{2\pi\ell_s^2} = \frac{\ut \, \rt_\infty}{2\pi\ell_s^2} \,.
\label{quarkmass}
\eeq
The quark condensate can now be computed using a simple argument.
The Hamiltonian density of the theory can be written as
\beq
\calh = \calh_0 + \mq \int \df^2\theta \, \tilde{Q}Q \,,
\eeq
where $\calh_0$ is $\mq$-independent, and $\tilde{Q}$, $Q$ represent
the hypermultiplet superfields in $\cN=1$ notation. It follows that
\beq
\frac{\delta \cale}{\delta \mq} =
%\frac{\delta}{\delta \mq} \langle \calh \rangle =
\langle  \int \df^2\theta \, \tilde{Q}Q \, \rangle = \cc \,,
\eeq
where $\cale = \langle \calh \rangle$ is the vacuum energy density. 
In the last equality, we have assumed that the vacuum expectation 
value of the fundamental scalars vanishes, since they are massive 
(even with $\mq=0$ after supersymmetry breaking), that is, the energy 
density increases if they acquire a non-zero expectation value.

Since in the string description the quark mass is given by the
asymptotic position of the D6-brane, we need to evaluate the change in
energy of the D6-brane associated to a change of the boundary
condition $\rt_\infty$. In view of (\ref{eq:D6action}) this is given by
\beq
\delta \cale = - \int \df\lambda \df\Omega_{\it 2} \, \delta \cL =
\t6 \ut^3 \int \df\lambda \df\Omega_{\it 2} \, \delta
\left[\sqrt{h} \left( 1 + \fc{1}{4\rho^3} \right)^2 \, \l^2 \sqrt{1 +
\dot{r}^2} \right] \,,
\eeq
where $\cL$ is the D6-brane Lagrangian density rescaled in accordance
with \reef{tildes}. Using the equations of motion this becomes
\beq
\delta \cale = -4\pi \left.\left(\d r\frac{\partial}{\partial\dot{r}}
\frac{\cL}{\sqrt{h}}\right)\right|^{\l=\infty}_{\l=0} =
4\pi \t6 \ut^3\left. \left(1+\frac{1}{4\rho^3}\right)^2
\lambda^2\frac{\dot{r}\, \d r}{\sqrt{1+\dot{r}^2}}
\right|^{\l=\infty}_{\l=0} =
- 4\pi \t6 \ut^3 \, c\, \delta r_\infty  \,,
\eeq
where we used \eqn{rt} and the fact that $\dot{r}|_{\l =0}=0$,
by rotational symmetry in the 567-directions. In view of
\eqn{quarkmass} we finally obtain
\beq
\cc = \frac{\delta \cale}{\delta m_\mt{q}} =
-8 \pi^2 \ell_s^2 \t6 \ut^2 \, c \,.
\label{quarkcond}
\eeq
As anticipated, the quark condensate is directly related to
$c$; we will see below that $c$ is determined by $\rt_\infty$.
Note that, in terms of gauge theory quantities,
\be \fc{1}{\nc} \cc \simeq \gym^2 \nc \mkk^3 c\,. 
\ee
The normalization on the left-hand side is that expected in the 't
Hooft limit.

We now return to the full equation of motion (\ref{resc}), which we
will solve numerically. In order to understand the nature of the 
solutions better, however, it is
helpful to first consider the limit $\rt_\infty \gg 1$, in which
(approximate) analytical solutions can be found.

If $\rt_\infty \gg 1$, the entire D6-brane should lie far
away from the `bolt' at the center of the \nonsol, hence we 
expect the solution to be a small perturbation around the solution
for $\ut=0$. In other words, we set $\rt(\lt)=\rt_\infty + \d \rt
(\lt)$ and assume that $\d \rt \ll 1$. Substituting this into
\eqn{resc} and expanding to linear order in $\d \rt$ we find
\beq
\frac{d}{d\lt}\left(\left. \lt \right. ^2 \delta\dot{\rt}\right) \simeq
-\frac{3}{2} \lt^2 \frac{\rt_\infty}{\left( \left. \lt \right.^2+ \left.
\rt \right._\infty^2 \right)^{5/2}} \,,
\eeq
which is easily integrated with the result
\beq
\d \dot{\rt} \simeq -\frac{3}{2}\frac{\rt_\infty}{\left. \lt
\right.^2} \int_0^{\lt} \df x\, \frac{x^2}
{\left( x^2+ \left. \rt \right._\infty^2 \right)^{5/2}} =
- \frac{1}{2 \rt_\infty} \, \frac{\lt}
{\left( \left. \lt \right. ^2 +
\left. \rt \right._\infty^2 \right)^{3/2}} \,.
\eeq
Here we have imposed the boundary condition $\dot{\rt} = 0$ at
$\lt=0$, as required by regularity of the solution at the origin of
the 567-space. Integrating once more, with the boundary condition
$\d \rt|_{\lt=\infty}=0$,  we find
\beq
\rt(\lt) \simeq \rt_\infty + \frac{1}{2 \rt_\infty} \,
\frac{1}{\sqrt{\left. \lt \right. ^2 + \left. \rt \right._\infty^2}} \,.
\label{asymptotic}
\eeq
From this large-$\rt_\infty$ solution, we see  that the
D6-brane bends `outwards', that is, it is `repelled' by the
D4-branes. This repulsion persists for arbitrary values
of $\rt_\infty$, as is confirmed by numerical
analysis. Eq.~\eqn{resc} can be integrated numerically for
any value of $\rt_\infty$, and we have plotted solutions for several
values of $\rt_\infty$ in Figure \ref{fig:D6embedding}.

\FIGURE{\epsfig{file=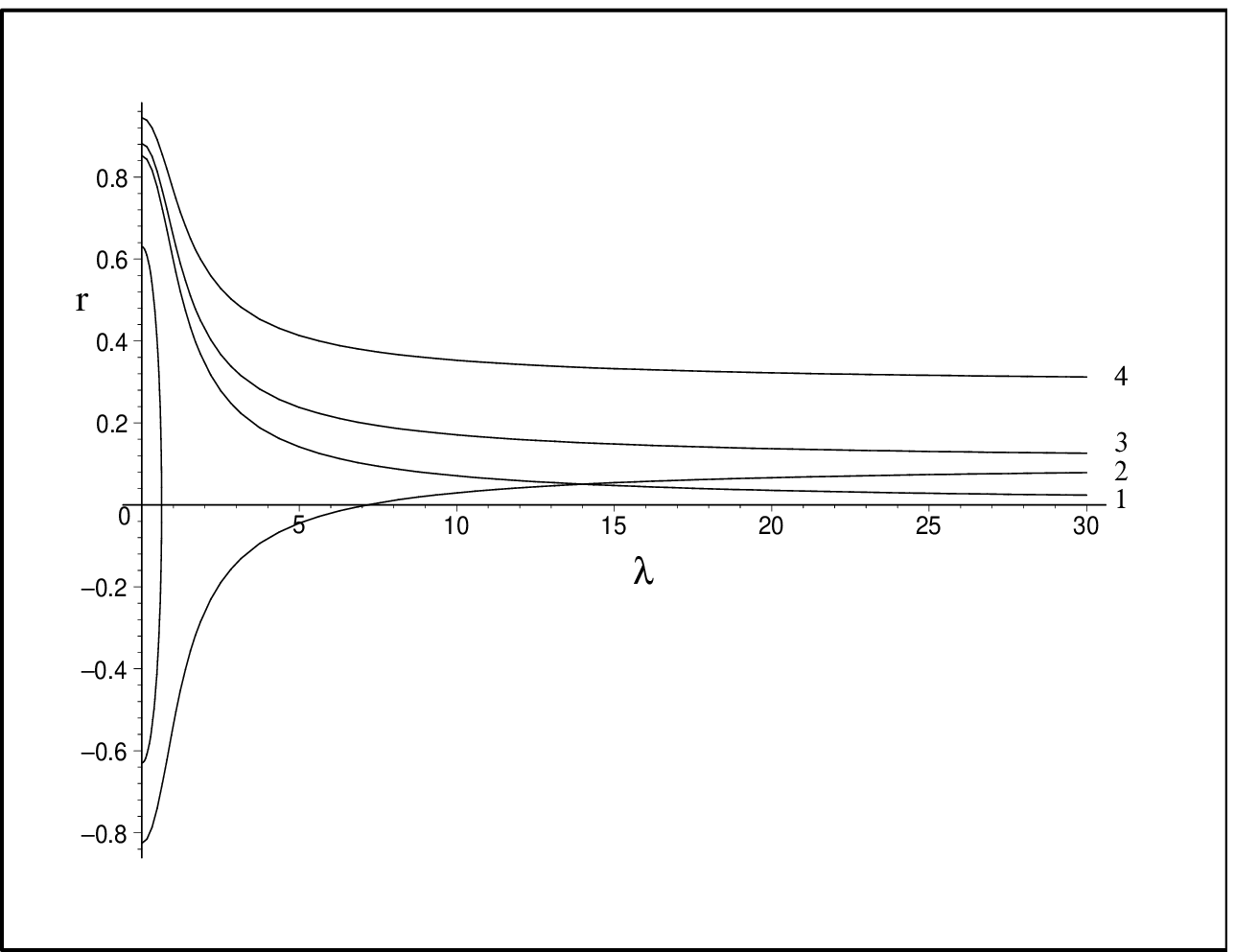, height=10cm}
\caption{The D6-brane embedding for several values of the quark
  mass $\mq \propto \rt_\infty$. The interior of the `circle' in the center is
  the region $U<\ut$, which is actually not part of the space.
  The `2' and `3' embeddings
  correspond to the same value of $\rt_\infty$ but have opposite-sign
  $c$'s; the `3' embedding has lower energy.}
\label{fig:D6embedding} }

Figure \ref{condensate} displays the function $c(\rt_\infty)$
found by numerical integrations. As mentioned above, the value of
$\rt_\infty$ together with the requirement of regularity at
$\lt=0$ (\ie, $\dot{\rt}|_{\lt=0}=0$) determines the solution
completely. Hence choosing $\rt_\infty$ fixes $c=c(\rt_\infty)$.
Recalling eqs.~\reef{quarkmass} and \reef{quarkcond}, which relate
$\rt_\infty$ and $c$ to the quark mass and condensate,
respectively, we see that this corresponds precisely to what is
expected on field-theoretic grounds: once the quark mass is specified, 
the infrared dynamics determines the chiral condensate.  In the field
theory we expect the $\ua$ symmetry to be spontaneously broken by a
non-zero condensate in the limit $\mq \ra 0$. This is indeed confirmed
by the numerical results for the D6-brane embedding, since we
see that, in figure \ref{condensate}, $c(\rt_\infty)$ approaches a
finite constant in the limit $\rt_\infty \ra 0$. In this limit,
the asymptotic boundary condition on the D6-brane respects the
rotational symmetry in the 89-plane, but it is energetically favourable
for the D6-brane to bend out in the 89-plane. 
Hence the embedding spontaneously breaks this rotational symmetry, 
which matches the $\ua$ symmetry-breaking in the
field theory.

We see from \eqn{asymptotic} that $c(\rt_\infty) \sim 1/ (2\rt_\infty)$
for large $\rt_\infty$, which is confirmed by the numerical
results. This implies that the chiral condensate scales as $1/\mq$ for
large $\mq$, as expected on field theory grounds \cite{SVZ79}.
A quick, somewhat heuristic argument for this in QCD is as
follows.\footnote{We thank D.\ T.\ Son for explaining this point to us.}   
The trace of the energy-momentum tensor reads 
\be
T^\mu_{\,\,\,\, \mu} = \mq \bar{\psi} \psi -
\fc{\alpha_s (11 \nc -2 \nf)}{24 \pi} \, \mbox{Tr} F^2 \,,
\ee
where we have explicitly written down the classical contribution, as
well as the anomalous one coming from the one-loop beta-function. 
In the limit $\mq \ra \infty$ the theory resembles QCD with no quarks,
for which 
\be
T^\mu_{\,\,\,\, \mu} = -\fc{11 \nc \alpha_s}{24 \pi} \,\mbox{Tr} F^2  \,.
\ee
Taking vacuum expectation values and equating the two expressions we find 
\be
\cc = \fc{\alpha_s \nf}{12 \pi \mq} \, \langle \mbox{Tr} F^2 \rangle \,.
\ee
Assuming that $\langle \mbox{Tr} F^2 \rangle \neq 0$ we conclude that 
$\cc \propto 1/\mq$.

For each value of $\rt_\infty>0$, the numerical analysis actually
finds two regular solutions, which have $c$'s of opposite sign. 
The curves labelled `2' and `3' in figure
\ref{fig:D6embedding} are a representative example of such a pair
of solutions.\footnote{ To describe the second solution (curve 2),
we allow $\rt(\lt)<0$ and restrict $\phi$ to the range $[0,\pi)$.}
We have confirmed that the solutions with positive $c$, for which
$\rt(\lt)>0$ for all $\lt$ (\eg, curve 3), have the lower energy
density. The negative-$c$ solutions bend around the opposite side of
the bolt at the center of the \nonsol\ geometry. Naively, these
solutions should be unstable, since the D6-brane is free to `slide
off the bolt' by moving out of the $\phi=\phi_0$ plane, and relax to the 
lower-energy, positive-$c$ solution. We will
return to this issue in the next section, where we study
fluctuations of the D6-brane worldvolume fields around embedding
solutions of each type --- in fact, we will find tachyonic
fluctuations for the negative-$c$ solutions, confirming the
intuition given above.

\FIGURE{\epsfig{file=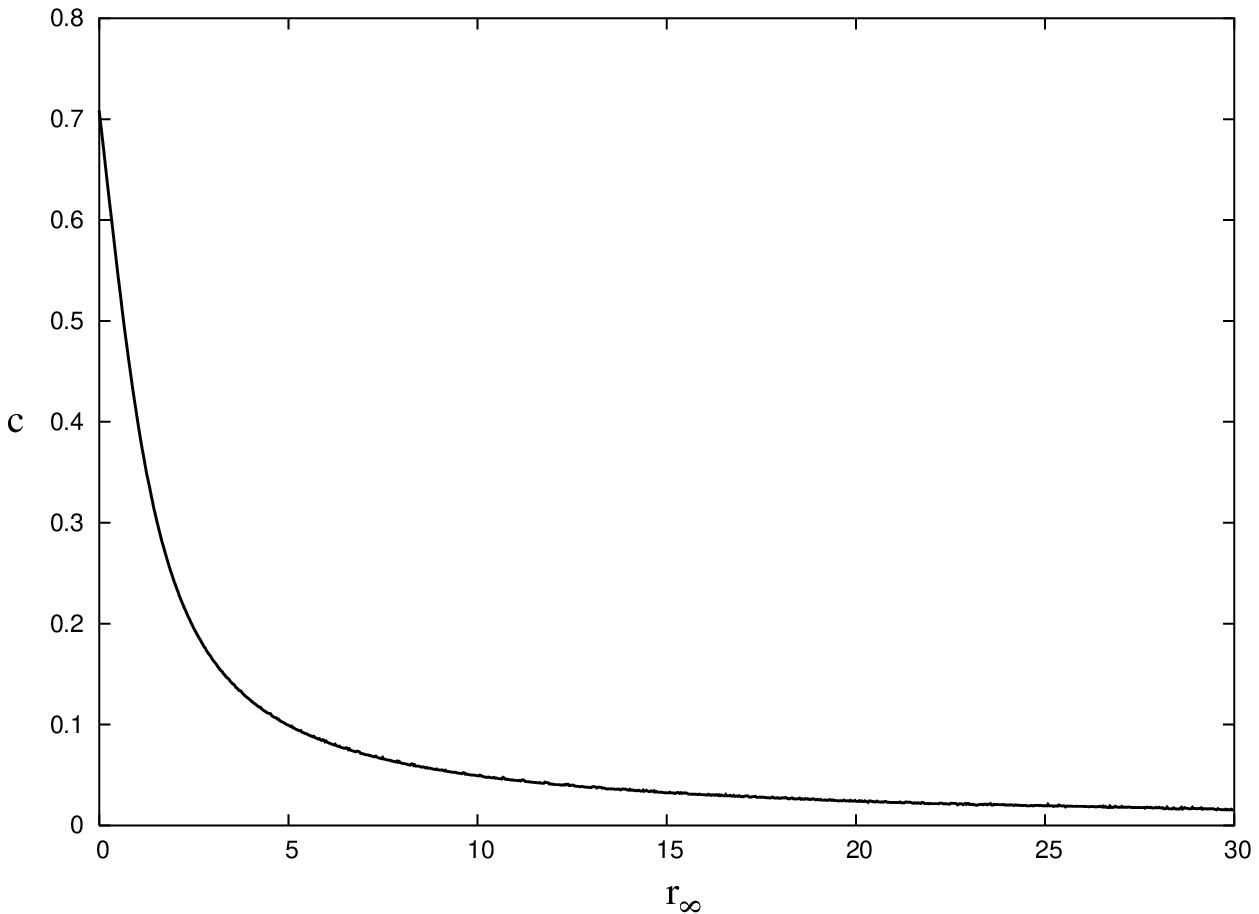, height=10cm}
 \caption{The quark condensate
   $\langle \bar{\psi} \psi \rangle \propto c$ as a function of the
   quark mass $\mq \propto \rt_\infty$.}
\label{condensate}}

Note that $r(\l)=0$ is an exact mathematical solution of
eq.~\eqn{embedeq} even if $\ut\neq 0$. Therefore one might be
tempted to conclude that it provides an alternative,
$\ua$-preserving solution whose energy should be shown to be
higher than that of the $\ua$-breaking solution above in order to
establish spontaneous chiral symmetry breaking. However, the
solution $r(\l)=0$ is not physically acceptable if $\ut \neq 0$
because it corresponds to an open D6-brane. That is, this solution
yields a D6-brane with a boundary, which is forbidden by charge
conservation. The boundary is easily seen as follows. With
$r(\l)=0$ we have $\rho =\l$, so this solution terminates at the origin
of the $U\tau$-plane, that is, at $U=\ut$. Since, at this
point, the $S^2$ in the metric \eqn{induced} still has a finite
radius, $R^{3/4}\ut^{1/4}$, the D6-brane would have an 
$\bbr{1,3} \times S^2$-boundary if it terminated at $U=\ut$. Note that if
$\ut=0$ then the two-sphere radius shrinks to zero-size and the
six-brane closes off at $U=0$. Hence the $\ua$-preserving
solution $r(\l)=0$ is physically sensible in the supersymmetric
case. This result is in agreement with the fact that unbroken
supersymmetry forbids a non-zero chiral condensate. Note, however,
that this is only a formal argument, since in the limit $\ut \ra 0$ the
curvature at $U=\ut$ diverges, meaning that the supergravity description
cannot be trusted in this region \cite{IMSY98}.

In section \ref{Dbar}, we will see that the solution $r(\l)=0$ can
play a role in the case $\ut \neq 0$ if it is simply extended past
the origin of the $U\tau$-plane. The physical interpretation of
such a solution is not that of a D4/D6 intersection, but rather
that of a D4/D6/$\overline{\mbox{D6}}$ intersection.

%%%%%%%%%%%%%%%%%%%%%%%%%%%%%%%%%%%%%%%%%%%%%%%%%%%%%%%%%%%%%%%%%%%%%%%%%%%%%%
%%%%%%%%%%%%%%%%%%%%%%%%%%%%%%%%%%%%%%%%%%%%%%%%%%%%%%%%%%%%%%%%%%%%%%%%%%%%%%
%%%%%%%%%%%%%%%%%%%%%%% SECTION  %%%%%%%%%%%%%%%%%%%%%%%%%%%%%%%%%%%%%%%%%%%%%
%%%%%%%%%%%%%%%%%%%%%%%%%%%%%%%%%%%%%%%%%%%%%%%%%%%%%%%%%%%%%%%%%%%%%%%%%%%%%%
%%%%%%%%%%%%%%%%%%%%%%%%%%%%%%%%%%%%%%%%%%%%%%%%%%%%%%%%%%%%%%%%%%%%%%%%%%%%%%
\section{Meson spectroscopy from D6-brane fluctuations ($\nf =1$)} 
\label{fluct}

Here, we consider a certain class of fluctuations of the
D6-brane around the embeddings described in the previous section.
We consider first the positive-$c$ (negative-condensate)
embeddings, which were argued to be stable and so should correspond
to the true `vacuum state' for a given quark mass. For simplicity
we restrict ourselves to fluctuations of the fields $r$ and
$\phi$, as these are sufficient to illustrate the physics we wish
to exhibit. We are therefore considering embeddings of the D6-brane 
of the form
\beq 
\phi=0+\d\phi \sac r=\rvac(\l)+\d r \sac \tau=\mbox{constant}
\,, \label{fluctansatz} 
\eeq
where the fluctuations $\d \phi$ and $\d r$ are functions of 
{\it all} of the worldvolume coordinates and $\rvac(\l)$ is the
numerically-determined vacuum embedding of the previous section.

In corroboration of our previous argument for the stability of
these embeddings, we find, in particular, that the spectra of
these fluctuations are non-tachyonic. We also briefly consider
fluctuations around the negative-$c$ (positive-condensate)
embeddings and show their spectra do contain tachyonic modes,
making manifest the instability of these configurations.

In the dual gauge theory, the $r$- and the $\phi$-fluctuations
correspond to a class of scalar and pseudo-scalar mesons, 
respectively. To see this, consider the complex scalar
field parametrising the 89-plane, $X = X_8 + i X_9 = r e^{i\phi}$. 
Gauge theory Lorentz transformations act only on the
0123-directions, under which $X$ is inert, so $X$-fluctuations 
clearly correspond to spin-zero mesons. However, a parity 
transformation in the gauge theory corresponds to a ten-dimensional
transformation that not only reverses the sign of (say) $X_3$, 
but also replaces $X$ by its complex conjugate, $\bar{X}$; this
leaves $r$ invariant but reverses the sign of $\phi$. 
From the gauge theory viewpoint, this is clear from the
presence in the microscopic Lagrangian of a coupling of the form 
$\psi_\mt{R}^\dagger X \psi_\mt{L} + 
\psi_\mt{L}^\dagger \bar{X} \psi_\mt{R}$: the transformation 
$X_3 \ra -X_3$ exchanges $\psi_\mt{R}$ and 
$\psi_\mt{L}$, so this term is only invariant if, simultaneously,
$X \ra \bar{X}$. From the string theory viewpoint, the fact that only
the combination of both transformations is a symmetry can be seen 
by examining the fermionic spectrum of the D4-D6 open strings.
This arises from the zero modes of the world-sheet fermions 
in the Ramond sector, that is, in the NN and DD directions: 
the 23-  and 89-directions (in the light-cone gauge) in 
\eqn{intersection}. Each of these modes is labelled by its weight
under rotations in each of the 23- and 89-planes, $(s_1, s_2), s_i=\pm 1/2$. 
The GSO projection requires $s_1=-s_2$. Since under $X_3 \ra -X_3$ we
have $(s_1, s_2) \ra (-s_1, s_2)$, and under $X_9 \ra -X_9$ we have 
$(s_1, s_2) \ra (s_1, -s_2)$, we see that the spectrum is only
invariant under the combination of both transformations.

%%%%%%%%%%%%%%%%%%%%%%%%%%%%%%%%%%%%%%%%%%%%%%%%%%%%%%%%%%%%%%%%%%%%%
\subsection{Analysis of the spectrum}
\label{anal}
%%%%%%%%%%%%%%%%%%%%%%%%%%%%%%%%%%%%%%%%%%%%%%%%%%%%%%%%%%%%%%%%%%%%%

Using the background metric \eqn{isometric} with the rescalings
\eqn{tildes}, the pullback of \eqn{isometric} to the embedding
given in \eqn{fluctansatz} is
\beqa
ds^2&=&\left(\frac{U}{R}\right)^{3/2}\eta_{\mu\nu}dx^\mu dx^\nu+
K[(1+\dot{r}_\mt{vac}^2)d\l^2 + \l^2d\O_{\it 2}^2+
2\dot{r}_\mt{vac}\pf_a(\d r) d\l dx^a]\non\\
&+&K[\pf_a(\d r) \pf_b(\d r) dx^a dx^b+ (r_\mt{vac}+\d
r)^2\pf_a(\d \phi) \pf_b(\d \phi) dx^a dx^b]\ , \label{indfluc}
\eeqa
where $a$ and $b$ run over {\it all} of the worldvolume
directions. To quadratic order in fluctuations, the D6-brane
Lagrangian density is then
\beqa \call =\call_0&-&\t6 \ut^3 \l^2 \sqrt{h}
\sqrt{1+\dot{r}_\mt{vac}^2}
\Bigg\{\left(\frac{3(7\r_\mt{vac}^2-\l^2)}{16\rho_\mt{vac}^{10}}+
\frac{3(4r_\mt{vac}^2-\l^2)}{4\rho_\mt{vac}^7}\right)(\d r)^2\non\\
&&\hspace{2cm}-\frac{3r_\mt{vac}}{4\rho_\mt{vac}^5}\left(1+
\frac{1}{4\rho_\mt{vac}^3}\right) \frac{\dot{r}_\mt{vac} \pf_\l(\d
r^2)}{1+\dot{r}_\mt{vac}^2}\non\\
&&+\left(1+\frac{1}{4\rho_\mt{vac}^3}\right)^2\, \sum_a
\frac{K}{2g_{aa}} \left(\frac{[\pf_a(\d
r)]^2}{1+\dot{r}_\mt{vac}^2}+ r_\mt{vac}^2[\pf_a(\d
\phi)]^2\right)\Bigg\}\ , \label{expandlag} \eeqa
where $\call_0$ is the Lagrangian density evaluated for the vacuum
embedding, as in eq.~\eqn{eq:D6action}. Here,
$\rho_\mt{vac}^2=\l^2+r_\mt{vac}^2$ and the $g_{aa}$ in the last
line correspond to the metric coefficients from eq.~\eqn{induced}.
In particular then, they contain no dependence on the
fluctuations. Of course, integration by parts and the equation of
motion \eqn{embedeq} for $r_\mt{vac}$ allowed the terms linear in
$\d r$ to be eliminated.

The linearised equations of motion that follow are, for $\d \phi$:
\beqa
&&\fc{9}{4\mkk^2} \frac{\rvac^2}{\rho_\mt{vac}^3}
\left(1+\frac{1}{4\rho_\mt{vac}^3}\right)^{4/3}\pf_\mu \pf^\mu (\d\phi)
+\frac{\rvac^2}{\l^2}
\left(1+\frac{1}{4\rho_\mt{vac}^3}\right)^2\nabla^2 (\d\phi) \non\\
&&\hspace{2cm}+\frac{1}{\l^2\sqrt{1+\dot{r}_\mt{vac}^2}}
\frac{d}{d\l}\left[\frac{\l^2 \rvac^2}{\sqrt{1+\dot{r}_\mt{vac}^2}}
\left(1+\frac{1}{4\rho_\mt{vac}^3}\right)^2\pf_\l (\d\phi)\right]=0 \,,
\label{phi}
\eeqa
and, for $\d r$:
\beqa
&&\frac{\l^2}{\sqrt{1+\dot{r}_\mt{vac}^2}}
\left[\fc{9}{4\mkk^2} \frac{1}{\rho_\mt{vac}^3}
\left(1+\frac{1}{4\rho_\mt{vac}^3}\right)^{4/3}\pf_\mu \pf^\mu (\d r)
+\frac{1}{\l^2}\left(
1+\frac{1}{4\rho_\mt{vac}^3}\right)^2\nabla^2 (\d r)\right] \non\\
&& \hspace{1cm}+\frac{d}{d\l}\left[\frac{\l^2}{(1+\dot{r}_\mt{vac}^2)^{3/2}}
\left(1+\frac{1}{4\rho_\mt{vac}^3}\right)^2\pf_\l(\d r)\right]
-\frac{d}{d\l}\left[\frac{3\l^2\rvac\dot{r}_\mt{vac}}
{2\rho_\mt{vac}^5\sqrt{1+\dot{r}_\mt{vac}^2}}
\left(1+\frac{1}{4\rho_\mt{vac}^3}\right)\right] \d r \non\\
&&\hspace{4cm}-\l^2\sqrt{1+\dot{r}_\mt{vac}^2}\left(\frac{3(7\rvac^2-\l^2)}
{8\rho_\mt{vac}^{10}}+\frac{3(4\rvac^2-\l^2)}{2\rho_\mt{vac}^7}\right)\d r =0 \,.
\label{r}
\eeqa
In these expressions, $\nabla^2$ is the Laplacian on the
two-sphere.

From the form of the equations of motion, it is clear that we may
separate variables to write
\beq 
\d \phi=\calp(\l) e^{i k_\phi \cdot x}\, Y_{\ell_\phi
m_\phi}(S^2) \sac \d r=\calr(\l) e^{i k_r \cdot x}\, Y_{\ell_r
m_r}(S^2) \,, 
\eeq
where $Y_{\ell m}$ are the conventional spherical harmonics. We
then look for normalizable solutions with definite $S^2$-angular
momentum $\ell_{\phi,r}$ and four-dimensional mass
$M_{\phi,r}^2=-k_{\phi,r}^2$. For the sake of simplicity in the
following, we only focus on the case $\ell_{\phi,r}=0$ and find
the lowest-mass modes for each fluctuation.

We employ the shooting method to solve the linearized equations
of motion. First we use the asymptotic form of the equations  for
$\l\to\infty$ and the requirement of normalizability to determine
the asymptotic boundary conditions for the fluctuations. Then we
vary the four-dimensional mass parameters until, by numerically
integrating the equation of motion, we find solutions that are
also regular at the origin. Recall\footnote{Also recall that we
are working with dimensionless variables $r$ and $\l$, rescaled as
in eq.~\eqn{tildes}.} that the asymptotic behaviour of the vacuum
profile is $\rvac\simeq r_\infty + c/\l$, where $r_\infty$ is
related to the quark mass by equation \eqn{quarkmass}. In this
asymptotic region we have that $\rho_\mt{vac}(\l) \simeq\l$, so
the equation of motion for $\d\phi$ becomes, approximately,
\beq
\frac{9M_\phi^2}{4\mkk^2}\frac{\rvac^2}{\l}(\d\phi)+
\frac{d}{d\l}\left[ \l^2\rvac^2\pf_\l(\d\phi)\right] = 0\,,
\eeq
and similarly that for $\d r$ becomes
\beq
\frac{9M_r^2}{4\mkk^2}\frac{1}{\l}(\d r)+\frac{d}{d\l}
\left[ \l^2 \pf_\l(\d r)\right] = 0\,.
\eeq
It follows
that the asymptotic behaviour of $\d r(\l)$ does not depend on
that of $\rvac(\l)$, whereas that of $\d \phi(\l)$ depends
crucially on whether or not $r_\infty=0$, that is, whether or not
the quarks are massless. We will now show that if $r_\infty=0$
then there exists a normalizable $\d \phi$ mode with $M_\phi=0$,
whereas if $r_\infty \neq 0$ there is a mass gap in the spectrum.

\FIGURE{\epsfig{file=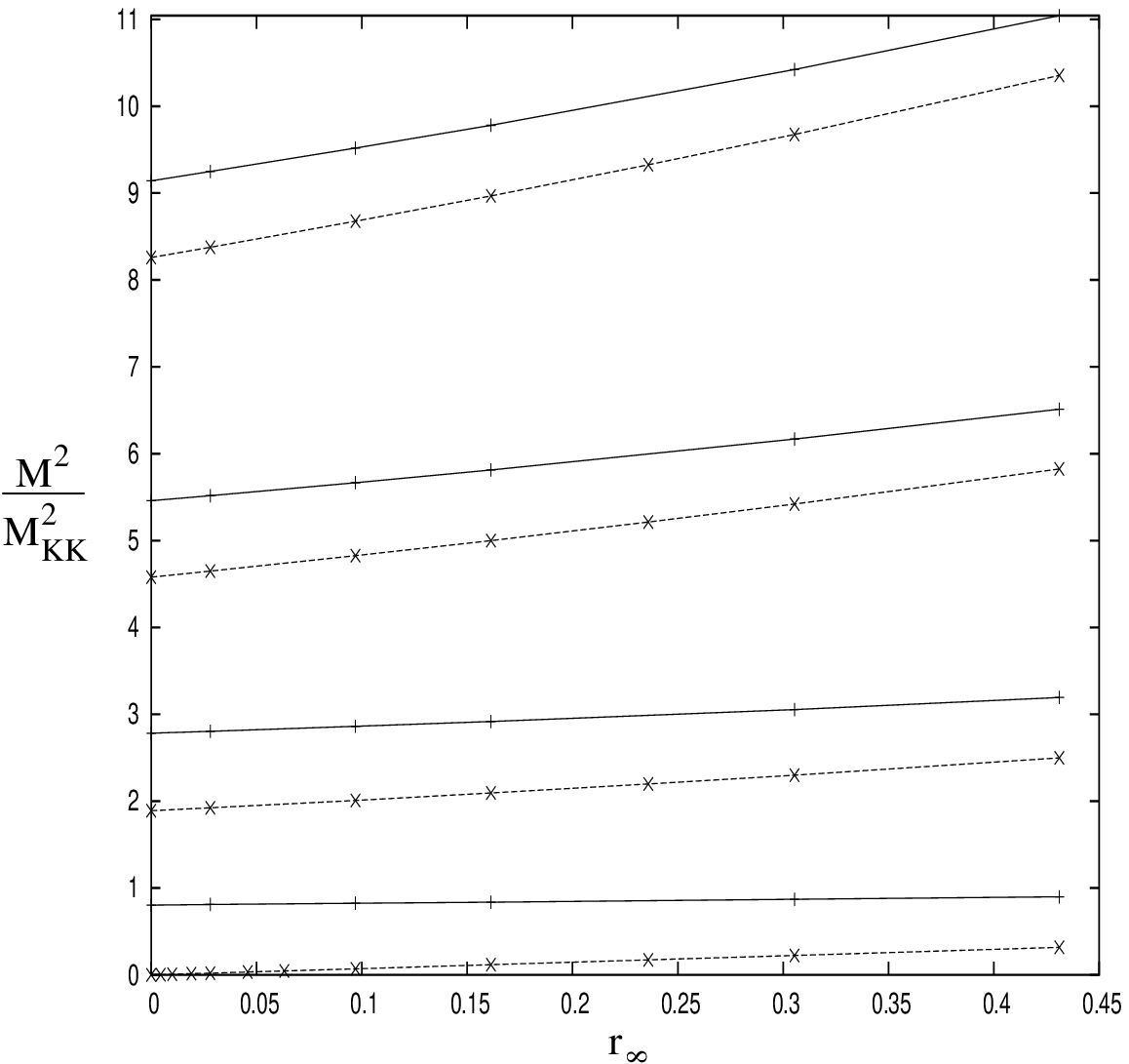, height=10cm} 
\caption{Squared masses for the four lowest-lying $\d\phi$ modes 
(dashed lines), dual to pseudo-scalar mesons, and the four 
lowest-lying $\d r$ modes (solid lines), dual to scalar mesons, 
as functions of the quark mass $\mq \propto r_\infty$.}  
\label{squaredscalarmasses} }

If we set $r_\infty=0$, we find that the asymptotic equation for
$\d \phi$ becomes
\beq
\frac{9M_\phi^2}{4\mkk^2}\frac{1}{\l^3}(\d\phi)+\pf_\l^2(\d\phi) = 0 \,.
\eeq
The first term is sub-leading for large $\l$, so we may drop it to
find that the asymptotic form of $\d \phi$, regardless of
the value of $M_\phi$, is
\beq
\d\phi\simeq b + a \l \,,
\label{m=0}
\eeq
where $a$ and $b$ are arbitrary constants. For $r_\infty\ne 0$,
however, we find that
\beq
\frac{9M_\phi^2}{4\mkk^2}\frac{1}{\l^3}(\d\phi)+\frac{2}{\l} \pf_\l (\d\phi) +
\pf_\l^2(\d\phi) = 0\,.
\eeq
Again the first term is sub-leading for large $\l$, so the solution is
\beq
\d\phi\simeq \tilde{a}+\tilde{b}/\l \,.
\label{mne0}
\eeq
Normalizable solutions correspond to the boundary conditions
$a=0$ and $\tilde{a}=0$ in the cases $r_\infty=0$ and $r_\infty \neq 0$,
respectively, since the terms with coefficients $b$ and $\tilde{b}$
are the ones that
fall off more rapidly at infinity. Since the differential equations we
are solving are linear, we may choose $b=\tilde{b}=1$ without loss of
generality, so we are left with the boundary conditions
$\{\d\phi=1 \ ,\ \d\phi'=0\}|_{\l\to\infty}$ if $r_\infty=0$, and
$\{\d\phi=1/\l\ ,\  \d\phi'=-1/\l^2\}|_{\l\to\infty}$ if $r_\infty\neq 0$.
A similar analysis for the $\d r$ fluctuations shows that they obey
the same asymptotic equation of motion as the $\d\phi$ fluctuations
in the $r_\infty\ne 0$ case, so the appropriate boundary conditions are
$\{\d r=1/\l\ ,\ \d r'=-1/\l^2\}|_{\l\to \infty}$.

\FIGURE{\epsfig{file=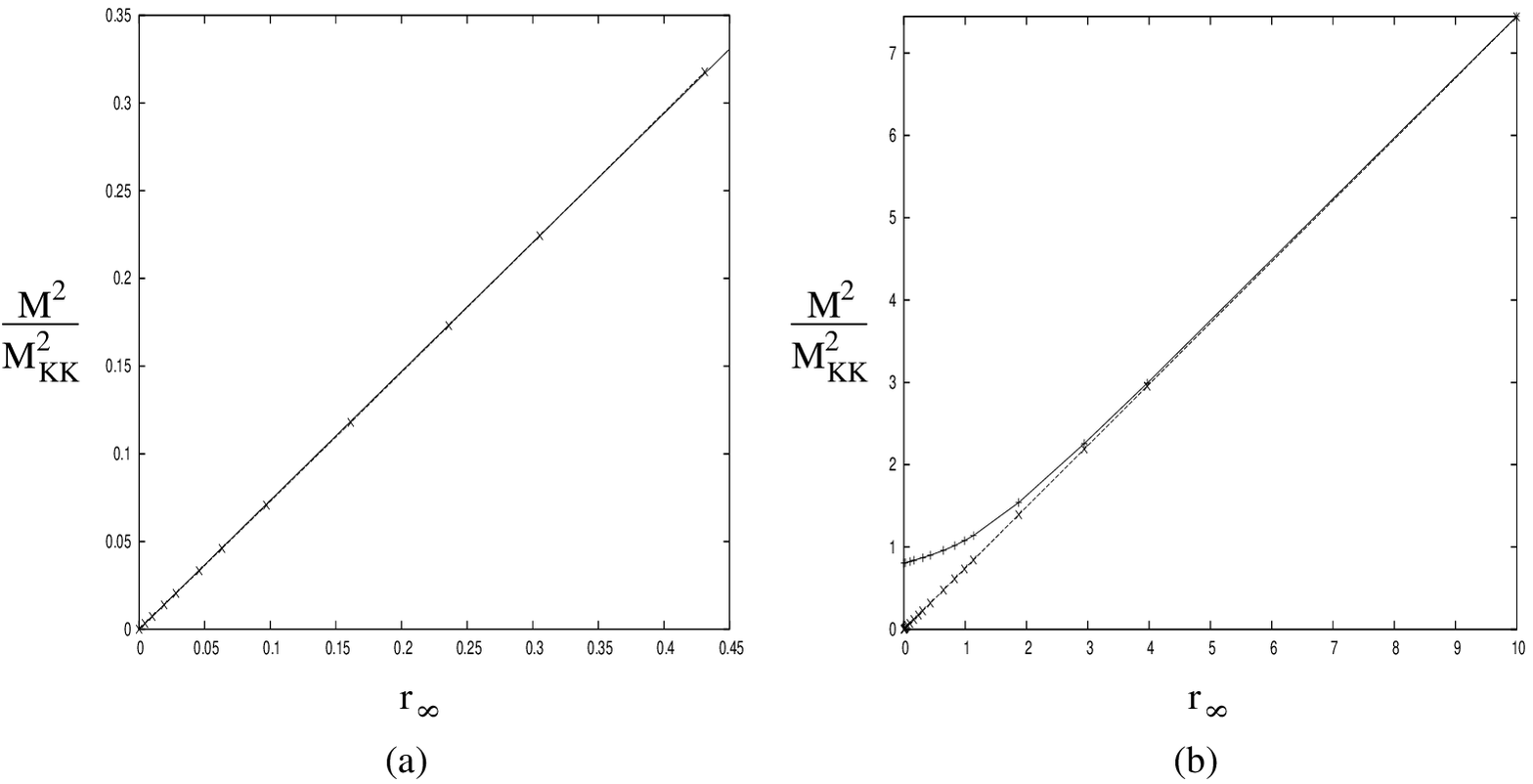, height=7cm}
\caption{(a) Linear fit of the lowest-lying $\d\phi$ mode 
  for different values of $\mq \sim \rt_\infty$.
  (b) Degeneracy of the lowest-lying $\d\phi$ (dashed) and $\d r$
  (solid) modes in the supersymmetric limit $r_\infty\gg 1$.}
\label{linearanddegenerate} }

Now note that $\d \phi = e^{i k_\phi \cdot x}$, with $k_\phi^2=0$,
is an exact, regular solution of \reef{phi} with zero
four-dimensional mass (and $\nabla^2(\d\phi)=0$), which is
normalisable {\it only} if $r_\infty=0$. This means that with
massless quarks there is a massless, pseudo-scalar meson in the 
spectrum of the
gauge theory, whose dual is the zero-mode fluctuation of the
D6-brane field $\phi$. The mathematical form of these
fluctuations shows that they correspond to rotations of the vacuum
embedding in the 89-plane. Of course, this form is precisely as
expected for the Goldstone mode associated with the spontaneous
breaking of this rotational symmetry.

The shooting technique allows us to verify this numerically, and
to find the values of $M_{\phi,r}$ for which other
normalizable, regular solutions exist. The results are the
low-lying (pseudo)scalar meson spectra, which are displayed in Figure
\ref{squaredscalarmasses} for different values of the quark mass.
As anticipated, there is no mass gap in the $\phi$-meson spectrum
when the quark is massless. There is a smooth transition to a
theory with a mass gap as the quark mass is increased and, in
fact, a best-fit (shown in figure \ref{linearanddegenerate}a) 
yields a linear relationship,\footnote{ 
An analogous relationship was observed in \cite{BEEGK03}.} 
$M_\phi^2 \simeq 0.73\, \mkk^2\, r_\infty$, for
small $r_\infty\propto m_\mt{q}$, as expected from the GMOR
relationship \eqn{GMOR}, which we will reproduce analytically 
in the next section.

As shown in figure \ref{squaredscalarmasses}, there are no
normalisable, regular solutions of equation \eqn{r} with
$M_{r} =0$, so there is a mass gap in the $r$-meson spectrum
regardless of the value of the quark mass. For small quark mass
the mass scale of all of the mesons, except the lowest-lying $\d
\phi$ mode, is\footnote{This also sets the scale of the glueball
mass spectrum \cite{COOT98}.}
\be
M^2 \sim \mkk^2 \sim \frac{U_\mt{KK}}{R^3}\,.
\label{region1}
\ee
This is another reflection of the lack of decoupling between the QCD
scale and the compactification scale.

As suggested by Figure \ref{squaredscalarmasses}, the $r$- and
$\phi$-mesons occur in pairs that become degenerate for large enough
a quark mass. The reason is that, in the supersymmetric case
($\ut=0$), these mesons belong to the same supermultiplet and hence have
equal masses, as in \cite{KMMW03}. If $\ut \neq 0$ this degeneracy is
only approximate for quark masses much larger than the
supersymmetry-breaking scale, $\msusy$, becoming exact in the limit
$\mq \ra \infty$. Geometrically, this corresponds to the fact
that the distortion of the D6-brane away from
the flat, supersymmetric profile decreases the farther away from the
D4-branes we embed it. Since this distance scale is measured by
$L \propto m_\mt{q}$, and $\ut$ characterises the supersymmetry
breaking, we expect that, for $L \gg \ut$, the characteristic
mass scale of the mesons should become independent of $\ut$ and
depend, instead, on some ratio of $L$ and $R$. To see that this is
indeed the case, we approximate the vacuum profile by the flat
solution in equations \reef{phi} and \reef{r}, that is, we set
$\rvac(\l) = r_\infty$, and we expand these equations assuming that
$r_\infty \gg 1$. In terms of a new variable $y= \l / L$ we obtain
(for zero $S^2$-angular momentum)
\beq
\frac{R^3 M_\phi^2}{L (y^2+1)^{3/2}} \d \phi +
\frac{1}{y^2}\frac{d}{dy}[y^2\pf_y\d\phi] = 0 \ ,
\label{phitrans}
\eeq
and
\beq
\frac{R^3 M_r^2}{L (y^2+1)^{3/2}} \d r +
\frac{1}{y^2}\frac{d}{dy}[y^2\pf_y\d r] = 0 \,.
\label{rtrans}
\eeq
The fact that the two types of fluctuations satisfy the same equation 
confirms that the spectrum becomes degenerate in this limit. It is
also clear that the mass scale for the mesons in this limit is now
\be
M^2 \sim \fc{L}{R^3} \sim \fc{\mq \mkk}{\gym^2 \nc} \,,
\label{region2}
\ee
since this is the only scale that appears in eq. \eqn{phitrans},
\eqn{rtrans}.

The transition between the regimes \eqn{region1} and \eqn{region2}
occurs at $L \sim \ut$, and is illustrated for the
lightest $r$- and $\phi$-mesons in figure \ref{linearanddegenerate}b, 
which was generated by solving equations \reef{phi} and \reef{r} 
numerically. The figure clearly illustrates the degeneracy of the
modes in the expected regime. The condition $L \sim \ut$
translates into $\mq \sim \msusy$, with $\msusy$ as defined in
\eqn{msusy}, so $\msusy$ is indeed the scale at which supersymmetry
is restored, as suggested by our choice of notation. Note that at the 
transition point, at which of course eqs. \eqn{region1} and \eqn{region2}
coincide, we have $M \sim \mq / (\gym^2 \nc)$. As we are working in 
the 't Hooft limit of the gauge theory, these mesons are very deeply bound, 
\ie, $M \ll 2\mq$. The suppression of these meson masses by the 
't Hooft coupling is even stronger than that found in a similar 
context in \cite{KMMW03}, where $M \sim \mq / \sqrt{\gym^2 \nc}$.
For large quark mass there is an extra suppression implicit in
eq. \eqn{region2} due to the fact that $M^2$ scales linearly in $\mq$,
as opposed to quadratically. 

As we are working in a range where we expect the gauge theory is
five-dimensional, it is perhaps more appropriate to re-express
the transition scale in terms of the five-dimensional gauge coupling;
it then reads $\mq \sim g_{\it 5}^2\nc \mkk^2$. Similarly, the mass scale
typical of the mesons becomes $M^2 \sim \mq/(g_{\it 5}^2\nc)$. In any
event, the transition scale suggests that the effective mass squared of
some of the microscopic degrees of freedom must be at least of the
order $\msusy \gg \mkk$, rather than just 
$\mkk$. Note that this mass scale is a factor of $\sqrt{\gym^2 \nc}$
larger than that coming from a simple one-loop calculation for the
scalar masses, which yields $\sqrt{\gym^2 \nc} \mkk$. 
This factor seems to be ubiquitous in extrapolations from weak to
strong 't Hooft coupling in the context of AdS/CFT.

As we saw above, the squared masses of all
mesons scale linearly in the quark mass as $\mq \ra \infty$. 
For the lightest meson, the results displayed in figure 
\ref{linearanddegenerate}b seem to indicate that, in fact, this linear 
relationship is valid not only in the limits of small and large $\mq$,
but for all quark masses. This result has been 
confirmed to within the best accuracy of our numerical analysis. 
Unfortunately, eq.~\eqn{phitrans} has four regular singular points 
and so no closed-form analytic solution is available, but its
numerical solution gives, for the lowest-mass modes,
\beq
M^2 \simeq 1.66\, \frac{L}{R^3} \simeq 
20.91\, \frac{\mkk}{\gym^2\nc} \mq \simeq 0.74\, \mkk^2 r_\infty\ .
\eeq 
The final form of this expression agrees with a best-fit of the data plotted
in the region of figure \ref{linearanddegenerate}b with $r_\infty\gg 1$ and, 
in principle, describes the extension of this figure to arbitrarily high values
of $r_\infty$. We therefore conclude that, in our model, the GMOR
linear relation between $M_\pi^2$ and $\mq$ extrapolates to all quark masses.

%%%%%%%%%%%%%%%%%%%%%%%%%%%%%%%%%%%%%%%%%%%%%%%%%%%%%%%%%%%%%%%%%%%%%
\subsection{Stability Issues}
\label{stable}
%%%%%%%%%%%%%%%%%%%%%%%%%%%%%%%%%%%%%%%%%%%%%%%%%%%%%%%%%%%%%%%%%%%%%

Recall that, at the end of section \ref{broken}, we remarked that
our numerical integrations had revealed two classes of six-brane
embeddings, those with positive $c$ and those with negative $c$.
From the numerical analysis, we could show that, for a given
$\rt_\infty$, the positive-$c$ or negative-condensate solutions
had the lower energy density. Further we argued that the
negative-$c$ solutions should be unstable.

The stability of the negative-condensate (positive-$c$) vacuum embedding is
reflected in the fact that the spectrum of fluctuations does not
contain tachyonic modes. While numerical searches did not reveal
any tachyonic fluctuations, one can formulate a general argument
that no such modes can exist in the spectrum of
$\d\phi$.\footnote{We expect that these arguments extend to $\d r$
fluctuations but we did not explicitly consider these modes a
possible decay channel between the two D6-brane embeddings, due to
the geometric obstruction provided by the `bolt' at the centre of
the \nonsol.} Consider the equation of motion \eqn{phi} (with zero
angular momentum on the two-sphere, for simplicity). Multiplying
by $\d\phi$, integrating over $\l$ and integrating by parts,
yields
\beqa &&\int_0^\infty\df \l\,
\left\{\frac{9M_\phi^2}{4\mkk^2}\frac{\l^2\rvac^2}{\rho_\mt{vac}^3}
\left(1+\frac{1}{4\rho_\mt{vac}^3}\right)^{4/3}
\sqrt{1+\dot{r}_\mt{vac}^2}\, \d\phi^2
-\frac{\l^2\rvac^2}{\sqrt{1+\dot{r}_\mt{vac}^2}}
\left(1+\frac{1}{4\rho_\mt{vac}^3}\right)^2
[\pf_\l(\d\phi)]^2\right\}\non\\
&&\hspace{4cm}=\left[\frac{\l^2\rvac^2}{\sqrt{1+\dot{r}_\mt{vac}^2}}
\left(1+\frac{1}{4\rho_\mt{vac}^3}\right)^2\d\phi\,
\pf_\l(\d\phi)\right]_0^\infty=0\ . \label{integrostability} \eeqa
Note that we were able to set the term on the right-hand side to
zero using the behaviour of the normalizable $\d\phi$ modes at
$\l=0$ and $\l=\infty$. On the left-hand side, the second term in
the integrand is manifestly negative or zero, while the sign of
the first term depends on that of $M_\phi^2$. Therefore, for
tachyonic modes it is clear that the integral will be negative-definite,  
which is inconsistent with the vanishing of the
right-hand side. The argument extends to include
nonvanishing $\ell_\phi$, which simply introduces an additional
negative-definite term under the integral. Hence we conclude that
the negative-condensate embeddings are free from tachyonic
instabilities in this sector. As this was the sector with the
lowest lying mode, we might have expected it to be the most
potentially problematic with regards to stability. Hence we are
confident that these positive-$c$ solutions are stable.

Now we turn our attention to the negative-$c$ or
positive-condensate embeddings. Recall that our intuition was that
these solutions should be unstable to `sliding off the bolt.'
Since the embedding coordinate ($r$, here) always vanishes at some
value of $\l$ for this case, we cannot address the issue of such
an instability using the $\d\phi$ equation of motion, since $\phi$
is not everywhere well-defined. 
Hence, we change to Cartesian
coordinates $X$ and $Y$ in the $r\phi$-plane. With the choice that
the D6-brane lies at $\phi_\mt{vac}(\l)=\phi_0=0$ (\ie,
$Y_\mt{vac}(\l)=0$), it follows that the embedding profile in $X$
satisfies the equation of motion \reef{resc} and we may set
$X_\mt{vac}(\l)=\rvac(\l)$. Fluctuations in $Y(\l)$ are now the
modes relevant for stability and have the benefit of being
well-defined for all $\l$. One finds the following equation of
motion for these fluctuations:
\beqa
&&\frac{9}{4\mkk^2}\frac{1}{\rho_\mt{vac}^3}
\left(1+\frac{1}{4\rho_\mt{vac}^3}\right)^{4/3}
\pf_\mu\pf^\mu(\d Y)
+\frac{1}{\l^2}\left(1+\frac{1}{4\rho_\mt{vac}^3}\right)^2
\nabla^2(\d Y)\non \\
&&+\frac{1}{\l^2\sqrt{1+\dot{X}_\mt{vac}^2}}\frac{d}{d\l}
\left[\frac{\l^2}{\sqrt{1+\dot{X}_\mt{vac}^2}}
\left(1+\frac{1}{4\rho_\mt{vac}^2}\right)^2\pf_\l(\d Y)\right]+
\frac{3(1+4\rho_\mt{vac}^3)}{8\rho_\mt{vac}^8}\d Y=0\ ,
\label{Y}
\eeqa
where, now, $\rho_\mt{vac}^2=\l^2+X_\mt{vac}^2$. Making a
stability argument along the lines of that given above for
$\d\phi$ is no longer possible, due to the presence of the final
term in \reef{Y}. This term introduces into the integrand
analogous to that in eq.~\reef{integrostability} a contribution that is
manifestly positive. Hence this approach does not yield a definite
conclusion in this case.
\FIGURE{{\epsfig{file=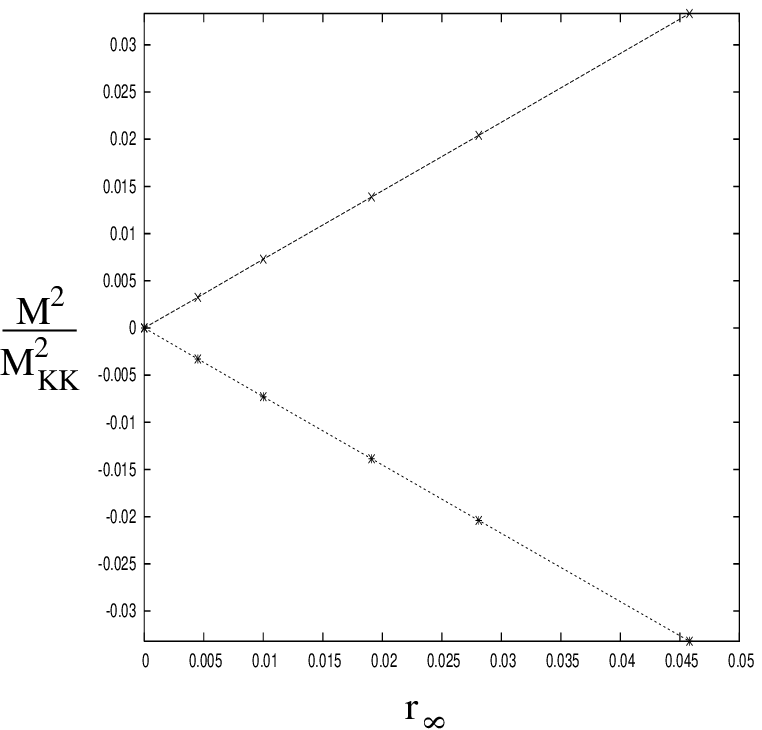, height=7cm}}
\caption{The lowest-lying $\d Y$ mode for some negative-condensate
embeddings (upper line), and the lowest-lying $\d Y$ mode for the
corresponding positive-condensate embeddings (lower line) --- the
$\d Y$ modes are tachyonic.}
\label{tachfig}}

However, it is still possible to look directly for tachyonic modes
in the spectrum of fluctuations around the negative-$c$ embeddings
using our numerical techniques. From eq.~\reef{Y} one can deduce
the asymptotic behaviour of the $\d Y$ modes to be
\beq 
\d Y\simeq \tilde{a}+\tilde{b}/\l\ , 
\eeq
independent of the vacuum embedding. Using the shooting method as
before, the spectrum of each positive-condensate embedding
considered numerically is found to contain a tachyonic mode and,
furthermore, these modes become `more tachyonic' as the quark mass
increases --- see figure \ref{tachfig}. As is suggested in the
figure, the squared mass of the lowest-lying $\d Y$ mode has a
linear dependence on the quark mass. In fact, a best-fit of the
data gives $M_Y^2\simeq -0.72\, \mkk^2 r_\infty$.

An examination of the next-to-lowest-lying $\d Y$ mode (not shown
in the figure) indicates that the mass of this mode, which is real
and positive for $m_\mt{q}=0$, also decreases as a function of
$r_\infty$. So for sufficiently large quark mass the spectrum may
contain more than one tachyon. However, we could not confirm this
directly because for large $r_\infty$ the negative-$c$ embeddings
approach extremely close to the bolt, and our computer code
behaved erratically in this situation.

%%%%%%%%%%%%%%%%%%%%%%%%%%%%%%%%%%%%%%%%%%%%%%%%%%%%%%%%%%%%%%%%%%%%%%%%%%%%%%
%%%%%%%%%%%%%%%%%%%%%%%%%%%%%%%%%%%%%%%%%%%%%%%%%%%%%%%%%%%%%%%%%%%%%%%%%%%%%%
%%%%%%%%%%%%%%%%%%%%%% SECTION  %%%%%%%%%%%%%%%%%%%%%%%%%%%%%%%%%%%%%%%%%%%%%
%%%%%%%%%%%%%%%%%%%%%%%%%%%%%%%%%%%%%%%%%%%%%%%%%%%%%%%%%%%%%%%%%%%%%%%%%%%%%%
%%%%%%%%%%%%%%%%%%%%%%%%%%%%%%%%%%%%%%%%%%%%%%%%%%%%%%%%%%%%%%%%%%%%%%%%%%%%%%
\section{The pseudo-Goldstone boson} \label{pseudo}

As discussed in the previous section, if $\mq=0$ the spectrum contains
a massless, pseudo-scalar mode that can be understood as the Goldstone boson
of the spontaneously broken $\ua$ symmetry. This is akin to the
$\eta'$ meson in QCD, which is massless in the large-$\nc$ limit. In an
abuse of language, however, we will refer to this Goldstone boson as a
`pion'. For non-zero quark mass, the $\ua$ symmetry is explicitly broken
and the pion becomes a pseudo-Goldstone boson with a mass $\mpi$
that, according to our numerical analysis, grows as $\mpi^2 \sim \mq$.
This means that at low energies, $E\ll \Lqcd$, the dynamics is
dominated by this particle. In this section we will provide an
analytic proof, using the string description, that the pion's mass
actually obeys the well-known Gell-Mann--Oakes--Renner relation
\eqn{GMOR}.

Recall that the gauge theory pion is dual to a $\phi$-fluctuation
of the D6-brane of the form $\delta \phi = e^{ik \cdot x} \vp (\l)$
that solves equation \eqn{phi} with $\nabla^2 (\d \phi)=0$, that is,
\beq
\frac{d}{d\l} \left[ p(\lambda)\dot{\vp} \right] =
-M^2 \mu(\lambda) \vp\ ,
\label{eq:eqvp}
\eeq
where $M^2 = -k^2$ is the four-dimensional mass,
\beqa
p(\lambda) &=& \lambda^2 \rvac^2
\left(1+\frac{1}{4\rho_\mt{vac}^3}\right)^2
\frac{1}{\sqrt{1+\dot{r}_{\mt{vac}}^2}} \,,
\label{pdef}
\\ \mu(\lambda) &=& \lambda^2 \frac{9}{4\mkk^2}\frac{\rvac^2}{\rho_\mt{vac}^3}
\left(1+\frac{1}{4\rho_\mt{vac}^3}\right)^{4/3}
\sqrt{1+\dot{r}_{\mt{vac}}^2} \,,
\label{pmu}
\eeqa
and $\rho_\mt{vac}^2 = \lambda^2 +\rvac^2$. 
The allowed solutions
are those that both satisfy $\dot{\vp}|_{\l =0} = 0$
and are normalizable with respect to the scalar product
\beq
\langle \vp_1 |  \vp_2 \rangle = \int_0^\infty \df\l\, \mu(\lambda)
\vp^*_1(\lambda) \vp_2(\lambda) \,.
\eeq
This defines an eigenvalue problem whose solution is
a complete set of orthogonal
eigenfunctions $\vp_n$ with eigenvalues $M^2_n$. Notice that
if $\vp(\l)= \mbox{constant}$ then the eigenvalue equation
is satisfied with $M^2=0$, but the mode is normalizable only if
$\rvac(\l) \ra 0$ as $\lambda\rightarrow\infty$.
If $\vp(\l)$ is not constant, multiplying eq.(\ref{eq:eqvp}) by $\vp$,
integrating both sides with respect to $\l$ and performing an integral by
parts in the left-hand side, we arrive at
\beq
M^2 = \frac{\int_0^{\infty} \df\l\, p(\l) \dot{\vp}^2}
{\int_0^{\infty} \df\l\, \mu(\l) \vp^2} >  0\ .
\eeq
Since from (\ref{pdef}) and (\ref{pmu}) we have $p(\l)>0$ and
$\mu(\l)>0$, it follows that $M^2$ is positive, as indicated.\footnote{
This is not valid when $\rvac$ vanishes at some point (which
is not the case studied here) --- see the more detailed discussion
after equation (\ref{integrostability}).}

Having said that, we now want to compute the mode with lowest
eigenvalue $M^2>0$ when $m_\mt{q}$ is non-vanishing but small, that
is, let us suppose that $\rvac\rightarrow r_\infty$ as
$\lambda\rightarrow\infty$, with $r_\infty$ small but non-zero. In
this case we expect that the normalizable eigenmode can be obtained
from the massless-quark zero-mode, namely $\vp(\l)=\mbox{constant}$,
by using perturbation theory. To this end, it is convenient to define
\beq
\psi(\lambda) = \sqrt{p(\lambda)} \, \vp (\l) \,,
\eeq
which satisfies the new eigenvalue equation
\beq
\ddot{\psi} - \frac{\ddot\Psi}{\Psi} \psi = -M^2\, \nu \, \psi \,,
\label{psieqn}
\eeq
where $\Psi=\sqrt{p}$, and $\nu=\mu/p$ is the new integration
measure. If $r_\infty=0$ then $\psi(\l)=\Psi(\l)$ is a normalizable
solution of \eqn{psieqn} with zero eigenvalue. Making a small change in
the boundary condition, $r_\infty\to\d r_\infty\gtrsim 0$, induces a small
change in the vacuum profile $\rvac(\l)$,
which in turn induces a small change in the functions $\Psi(\l)$ and
$\nu(\l)$ appearing in equation \eqn{psieqn}. Standard
quantum mechanics perturbation theory can then be applied to obtain the
new lowest eigenvalue.

Let us denote with a `bar' the quantities corresponding to
$r_\infty=0$, so that for $r_\infty\neq 0$ we have
 \beqa
\Psi &=& \bar{\Psi} + \delta \Psi \,, \\
\nu &=& \bar{\nu} + \delta \nu \,,
\eeqa
where $\delta \Psi$ and $\delta \nu$ are the differences induced
by the change in $\rvac(\l)$. The new lowest eigenvalue and its
associated wave-function, $M^2 \gtrsim 0$ and
$\psi =\bar{\Psi} + \delta \psi$, obey, to leading order,
\beq
\delta \ddot\psi -\frac{\ddot{\bar{\Psi}}}{\bar{\Psi}} \, \delta \psi
-\delta\left(\frac{\ddot\Psi}{\Psi}\right) \bar{\Psi} =
- M^2 \, \bar{\nu} \, \bar{\Psi} \,.
\eeq
Decomposing $\delta \psi $ as a linear combination
of massless-quark eigenfunctions,
$\delta\psi = \sum_{n=0}^\infty \alpha_n \bar{\psi}_n $, we obtain
\beq
\sum_{n=1}^\infty \alpha_n \bar{M}^2_n \bar{\nu} \bar{\psi}_n +
\delta\left(\frac{\ddot\Psi}{\Psi}\right) \bar{\Psi} =
M^2 \, \bar{\nu} \, \bar{\Psi} \,,
\eeq
where we have dropped the first term in the series, using
$\bar{M}_0=0$. Multiplying by $\bar{\Psi}$, integrating over
$\lambda$ and using the orthogonality condition $\int_0^{\infty}
\df\l\, \bar\nu \bar{\Psi} \bar{\psi}_n =0$ for $n>0$ (since
$\bar{\Psi}$ is the zero-mode eigenfunction and is real) we deduce
that
\beq
M^2 \int_0^{\infty} \df\l \, \bar{\nu}\, \bar{\Psi}^2 =
\int_0^\infty \df\l \, 
\delta\left(\frac{\ddot\Psi}{\Psi}\right) \bar{\Psi}^2 =
\int_0^\infty \df\l \, 
(\bar{\Psi}\delta\ddot{\Psi} -\ddot{\bar{\Psi}}\delta\Psi) =
\left[ \bar{\Psi} \delta \dot\Psi - \dot{\bar{\Psi}} \delta \Psi
\right]_{\l=0}^{\l=\infty} \,.
\label{m2}
\eeq
Recalling that
\be
\bar{\Psi} = \lambda \bar{r}_\mt{vac}
\left(1+\frac{1}{4\bar{\rho}_\mt{vac}^3}\right)
(1+\dot{\bar{r}}_\mt{vac}^2)^{-1/4} \,,
\ee
which implies
\bea
\delta \Psi &=& \lambda \, \delta r_\mt{vac}
\left(1+\frac{1}{4\bar{\rho}_\mt{vac}^3}\right)
(1+\dot{\bar{r}}_\mt{vac}^2)^{-1/4} -
\frac{3\lambda \bar{r}^2_{\mt{vac}}}{4\bar{\rho}_\mt{vac}^5}
(1+\dot{\bar{r}}_\mt{vac}^2)^{-1/4} \, \delta \rvac - \nonumber \\
&& - \l\, \bar{r}_\mt{vac} \left(1+\frac{1}{4\bar{\rho}_\mt{vac}^3}\right)
\frac{1}{2}(1+\dot{\bar{r}}_\mt{vac}^2)^{-5/4} \dot{\bar{r}}_\mt{vac} \,
\delta\dot{r}_\mt{vac} \,,
\eea
we see that the only contribution to the last term in equation
\eqn{m2} comes from $\l=\infty$. This is easily evaluated, since for
large $\l$ we have
\be
\bar{r}_\mt{vac} \simeq \frac{\bar{c}}{\lambda} +
\calo \left( \frac{1}{\lambda^2}\right) \sac
\bar{\Psi} \simeq \l\bar{r}_\mt{vac} \simeq \bar{c} +
\calo \left( \frac{1}{\l}\right) \,.
\ee
It follows that the new eigenvalue is given by
\beq
M^2 = \frac{\bar{c}\, \d r_\infty}
{\int_0^\infty \df\l\, \bar{\nu} \bar{\Psi}^2} =
\frac{\bar{c} \, \d r_\infty}{\int_0^\infty \df\l\, \bar{\mu}} \,.
\label{GMORsugra}
\eeq
To show that this formula is precisely the GMOR relation
\eqn{GMOR}, we need to identify the pion decay constant in the
string description. In this description, a shift of the chiral
angle that parametrizes the space of vacua related to each other
by the $\ua$ symmetry corresponds to a rigid rotation of the
D6-brane field $\phi$ (both take values between 0 and $2\pi$).
Similarly, the pion field is identified with the normalizable,
massless mode $\d\phi$, described in section \ref{anal}. The pion
decay constant can be read off from the normalization of the
kinetic term in the four-dimensional low-energy effective
Lagrangian for this mode. That is, we integrate the $\d\phi$-terms
in eq.~\eqn{expandlag} over $\l$ and the coordinates on $S^2$ and
as a result find a four-dimensional action of the form
\be S =  -\fc{\fpi^2}{2}\int \df^4x\, \pf_\mu (\d{\phi})
\pf^\mu(\d{\phi})\ , \ee
where $f_\pi$ is the pion decay constant. From
eq.~\reef{expandlag}, then we have
 \beqa
 \fpi^2 &=& \t6 \ut^3
\int\df\Omega_{\it 2}\, \sqrt{h} \int_0^\infty \df\l\,
\l^2\frac{9}{4\mkk^2}\frac{\bar{r}^2_\mt{vac}}{\bar{\rho}^3_\mt{vac}}
\left(1+\frac{1}{4\bar{\rho}^3_\mt{vac}}\right)^{4/3}
\sqrt{1+\dot{\bar{r}}^2_\mt{vac}}\non\\
& =& 4\pi \t6 \ut^3 \int_0^\infty
\df\l \, \bar{\mu}(\l)\ . \label{fpi} 
\eeqa
Combining this result with eqs.~\eqn{quarkmass} and
\eqn{quarkcond}, we see that eq.~\eqn{GMORsugra} is precisely the
GMOR relation \eqn{GMOR}, as anticipated.

We close this section by verifying the agreement between the
analytical expression \eqn{GMORsugra} above and the results obtained 
purely by numerical methods and displayed in figure 
\ref{linearanddegenerate}a. The integral in equation \eqn{GMORsugra} 
can be evaluated numerically with the result
\beq
\int_0^\infty \df\l \, \bar{\mu}(\l) = \frac{9}{4\mkk^2} \int_0^\infty \df\l \,
\frac{\lambda^2\bar{r}_\mt{vac}^2}{\bar{\rho}_\mt{vac}^3}
\left(1+\frac{1}{4\bar{\rho}_\mt{vac}^3}\right)^{4/3}
\sqrt{1+\dot{\bar{r}}_\mt{vac}^2} \simeq \frac{0.97}{\mkk^2} \,.
\label{muint}
\eeq
Similarly, the numerical result for the constant $c$ for zero
quark mass (see Figure \ref{condensate}) is $\bar{c} \simeq 0.71$.
Substituting these two values in equation \eqn{GMORsugra} we get
$M^2 \simeq 0.73\, \mkk^2 \d\rt_\infty$, in exact agreement with a best-fit
of the points in figure \ref{linearanddegenerate}a.

Recall from section \ref{broken} that for nonvanishing quark mass
we found that there are two possible values of the quark
condensate, differing in sign, but that only the negative condensate is
stable. In section \ref{stable} the positive-condensate solution
was explicitly shown to be unstable by showing that the
pseudo-Goldstone mode $\d Y$ was tachyonic for this configuration.
Since the field-theoretic arguments establishing the GMOR relation
apply regardless of the sign of the quark condensate, the squared 
mass of this mode should also obey the GMOR relation, at least
approximately, with an appropriate sign change. Recall that 
eq. \reef{GMOR} represents the first term in an expansion
around $\mq=0$. Now in the string description at $\mq=0$, the 
D6-brane configurations for positive and negative $c$ are
identical, except for a rotation of $\phi$ by $\pi$. Hence up to a
crucial sign, the numerical coefficients in eq.~\reef{GMORsugra}
are unchanged and hence the tachyonic mode should satisfy
$M_Y^2 \simeq -0.73\, \mkk^2 \d r_\infty$. From Figure
\ref{tachfig}, it does indeed appear that the slopes of the lowest-lying 
$\d\phi$ and $\d Y$ modes are equal, but opposite in sign. More precisely, 
our best-fit of the numerical results for negative $c$ gave 
$M_Y^2\simeq -0.72\, \mkk^2 r_\infty$, which gives agreement to within 
about 1.4\%.

%%%%%%%%%%%%%%%%%%%%%%%%%%%%%%%%%%%%%%%%%%%%%%%%%%%%%%%%%%%%%%%%%%%%%%%%%%%%%%
%%%%%%%%%%%%%%%%%%%%%%%%%%%%%%%%%%%%%%%%%%%%%%%%%%%%%%%%%%%%%%%%%%%%%%%%%%%%%%
%%%%%%%%%%%%%%%%%%%%%% SECTION  %%%%%%%%%%%%%%%%%%%%%%%%%%%%%%%%%%%%%%%%%%%%%%
%%%%%%%%%%%%%%%%%%%%%%%%%%%%%%%%%%%%%%%%%%%%%%%%%%%%%%%%%%%%%%%%%%%%%%%%%%%%%%
%%%%%%%%%%%%%%%%%%%%%%%%%%%%%%%%%%%%%%%%%%%%%%%%%%%%%%%%%%%%%%%%%%%%%%%%%%%%%%
\section{Multiple flavours ($\nf >1$) and a holographic Vafa--Witten theorem}
\label{vafa}

Consider a vector-like gauge theory with vanishing $\theta$-angle
(like QCD) and $\nf$ flavours, all with identical mass $\mq>0$.
The Vafa--Witten theorem states that the vector-like $U(\nf)$
flavour symmetry cannot be spontaneously broken \cite{VW84}. This
theorem does not exclude the possibility that, at $\mq =0$, the
$U(\nf)$-preserving vacuum becomes degenerate with one or more
$U(\nf)$-breaking vacua. In this section we discuss the
holographic realization of the Vafa--Witten theorem, within the
approximations implied by the validity of the
supergravity/Born--Infeld theory. In particular, we are
considering $\nc \ra \infty$ and $\nf \ll \nc$. Note that,
strictly speaking, the Vafa--Witten theorem need not apply to the
gauge theory on the D4/D6-brane system because of the presence of
Yukawa-like couplings between the fundamental fermions and the
adjoint scalars. However, the latter scalar fields are massive and
the theorem should certainly apply in a limit where the scalar
masses are sent to infinity. Further, one expects the same result
for large but finite masses since the vacuum should still not be
sensitive to the scalars. In the present case, the scalar masses
are likely of order $\mkk \sim \Lqcd$ and so these couplings would
not be suppressed. Hence, from the gauge theory viewpoint, 
it is not clear a priori whether the effect of these fields can 
change the vacuum structure to the extent of violating the Vafa--Witten
theorem.\footnote{We thank A.\ Nelson for a discussion on this
point.} The results of this section answer this question
in the negative.

The $U(\nf)$ global symmetry of the gauge theory corresponds, in
the string description, to the $U(\nf)$ gauge symmetry on the
worldvolume of $\nf$ D6-brane probes in the geometry \eqn{metric}.
Their dynamics is described by the so-called non-Abelian
Dirac--Born--Infeld (DBI) action \cite{Myers99}. This is a
generalisation of the single-brane Abelian theory
\eqn{eq:D6action}, in which both the gauge fields and the
transverse scalars are promoted to Hermitian, $U(\nf)$ matrices.
This means, in particular, that the positions of the D6-branes are
no longer commuting quantities in general. In the particular cases
in which the transverse scalar matrices can be simultaneously
diagonalised, each of their $\nf$ eigenvalues can be interpreted
as the position of one of the D6-branes along the corresponding
transverse direction.

For the $U(\nf)$ global symmetry to be realized in the gauge
theory, the mass matrix for the fundamental quarks must be
diagonal with all eigenvalues equal to $\mq$. In our string
picture, this translates into the boundary condition that,
asymptotically, all the D6-branes should have the same asymptotic
boundary conditions. That is, they are all aligned with each other
in the ($z^5,\ldots,z^9$) subspace and all lie at a distance 
$2\pi \ell_s^2 \mq$ from the origin. We would like to show that
the configuration (which obviously satisfies the boundary
condition above) in which all the D6-branes are exactly
coincident, and are embedded exactly as in the case of a single
D6-brane, is the minimum-energy solution of the equations of
motion that follow from the non-Abelian DBI action.

To avoid problems with the coordinate singularity at $r=0$, we
adopt the Cartesian coordinates $X$ and $Y$ introduced in section
\ref{stable} to describe the transverse positions of the D6-branes
in the $r\phi$-plane. Then the configuration for the $\nf$
coincident six-branes may be written as:
\be 
X(\l) = \rvac (\l) \cdot \bbi{} \sac Y(\l) = 0 \sac \tau =
\tau_0 \cdot \bbi{} \,, 
\label{conf} 
\ee
where $\rvac (\l)$ is the profile for a single D6-brane determined
in section 2 and $\bbi{}$ is the $\nf \times \nf$ identity matrix.
Since all the transverse scalars commute, there is no ambiguity in the
interpretation of this solution as $\nf$ overlapping D6-branes
lying along the curve $X(\l) = \rvac (\l)$.

Let us first argue that this configuration is a stable solution of
the non-Abelian DBI equations of motion. We start by expanding the
non-Abelian DBI action to quadratic order in fluctuations around
the above configuration. Since there is an overall single trace in
front of the action, to linear order only variations proportional
to the identity generator couple to the $U(1)$ background fields,
as given above. Hence the resulting expression for these fields is
exactly that obtained from the linear variation of the action for a
single D6-brane. It thus follows that the configuration \eqn{conf}
is a solution of the non-Abelian equations provided $\rvac (\l)$
solves the corresponding Abelian equations.

We further observe that, to quadratic order in fluctuations, the
non-Abelian DBI action reduces to the sum of $\nf^2$ identical
copies of the Abelian action, one for each of the $\nf^2$
independent fluctuations. The reasons for this are two-fold:
First, given that the background fields in \eqn{conf} are
proportional to the identity, the quadratic fluctuations must
again contribute a term in the overall $U(1)$. Second, the
difference between the non-Abelian and the Abelian DBI actions
consists entirely of terms that involve commutators of fields.
Hence, with the given background, the only possible nonvanishing
commutators will involve two fluctuations, but such expressions
are necessarily $SU(\nf)$-valued, \ie, traceless. It follows that 
upon tracing over the gauge indices (and integrating by parts), the
quadratic action will take the form
\be 
\int \df^7\s\,\sum_{\alpha=1}^{\nf^2} \d Z^{i\a}\, {\cal O}_{ij}\,
\d Z^{j\a}\ , 
\label{house} 
\ee
where $Z^i=X, Y$, $\a$ is the gauge index and ${\cal O}_{ij}$ 
is the same operator as appears in the
quadratic expansion of the Abelian action. Thus the linearized
equations of motion, and hence the spectrum, are just $\nf^2$ copies
of those in the Abelian theory describing a single D6-brane
embedding. It follows that the solution \eqn{conf} is stable
provided its Abelian analogue is stable, which we know to be the
case from the analysis of section \ref{fluct}.

Note that these $\nf^2$ fluctuations of the D6-branes, which give
rise to the $\nf^2$ copies of the spectrum, are distinct
physical excitations and {\it not} gauge-equivalent: each of them
is associated to a different generator of $U(N_f)$, so there is no
small $U(N_f)$ gauge transformation (that is, one that tends to
the identity at infinity in the 0123-directions) that takes one
mode into another. In particular, this implies that in the limit
$\mq \ra 0$ there are $\nf^2$ massless, pseudo-scalar particles! 
We will come back to this point in the last section.

The arguments presented so far show that \eqn{conf} provides a stable
extremum of the non-Abelian DBI action, and therefore a local minimum
of the energy of the $\nf$ D6-branes. Strictly speaking, this
does not exclude the possibility that there exist other minima. If this was
the case, then the true vacuum (or vacua) would be that (or those)
with the lowest energy. On physical grounds, however, the existence of
these other vacua is unlikely. The reason is as follows. If we ignore the
interactions of the D6-branes with each other, then it is clear that
\eqn{conf} is the minimum-energy solution amongst those satisfying the
boundary conditions described above, since in the absence of
interactions each D6-brane must minimise its own energy. If we
include open string tree-level interactions by describing the
D6-branes' dynamics with the non-Abelian DBI action,
then the only obvious configuration that satisfies the appropriate
boundary conditions is \eqn{conf}, which we have shown to
be a stable solution. Inclusion of one-loop (or higher) open string
interactions between the D6-branes (such as exchange of closed
strings) would presumably just strengthen the conclusion that
\eqn{conf} is the preferred solution. To see this, consider the simple
case of $\nf=2$ D6-branes that overlap with each
other asymptotically, but that are slightly misaligned in some
small region. Locally, this situation is analogous to that of two
planar D6-branes in flat space that are almost but not exactly
parallel, in which we know that there is a net force that tends to
align the branes. It seems plausible that a similar force
will act on the branes in the situation of interest here, and
therefore that the preferred configuration will be \eqn{conf},
in which all the branes are perfectly aligned with each other.

This conclusion provides a holographic realisation of the
Vafa--Witten theorem, since the $U(\nf)$ gauge symmetry on the
D6-branes is unbroken if and only if they are all exactly
coincident.

By continuity, the arguments above imply that, in the limit $\mq
\ra 0$, the $U(\nf)$-preserving configuration has an energy no
greater than that of any other configuration. Just like in the
Vafa--Witten theorem, however, this does not rule out the
possibility that in this limit some other state becomes degenerate
with that in eq.~\eqn{conf}. In fact, the appearance of
$\nf^2$ massless modes in the linearized spectrum at $\mq=0$ suggests 
that the D6-branes might be free to rotate independently in the
89-plane.\footnote{Note that this is not
  guaranteed by the existence of massless modes, since this does not 
  exclude the appearance of higher order potential terms for the 
  fluctuations, \eg\ $\delta Z^4$.}
To see that this is indeed the case, consider, for simplicity, the 
$\nf=2$ configuration 
\be 
X(\l) = \rvac (\l) \cos\phi_0 \cdot \bbi{} \sac  
Y(\l) = \rvac (\l)
\sin\phi_0 \cdot \sigma_{\it 3} \sac \tau = \tau_0 \cdot \bbi{}
\,, 
\label{confbis} 
\ee
where $\sigma_{\it 3} = \mbox{diag} (1, -1)$ and $\phi_0$ is some 
fixed angle. The radial vacuum profile above is that 
corresponding to $r_\infty=0$. This configuration consists of two 
D6-branes with the same radial profiles but separated an angle 
$2\phi_0$ in the 89-plane, and obviously satisfies the boundary
condition that the branes overlap at $\l \ra \infty$. To see that it solves
the non-Abelain equations of motion, consider expanding the 
non-Abelian DBI action to linear order in fluctuations around \eqn{confbis},
as before. 
Since commutator terms are irrelevant at this order, and $X$ and
$Y$ are diagonal, the non-Abelian action again reduces to the sum of
two identical Abelian actions. In fact, the same argument shows that not
only is the configuration \eqn{confbis} a solution of the non-Abelian 
equations, but also that its energy is, for all $\phi_0$, equal to
twice that of a single D6-brane. It follows that \eqn{confbis} is
stable for all $\phi_0$. To see this, suppose that it is
not, \ie, that there is some fluctuation around it that is tachyonic.
Presumably, this would mean that this solution can decay to another
one of lower energy. However, the latter would also have lower energy
than the solution with $\phi_0=0$, in contradiction with our
continuity argument above. 

Note that the previous reasoning applies equally well to any
distribution of $\nf$ D6-branes around the $\phi$-circle, since it
only relies on $X$ and $Y$ being diagonal. Thus, at $\mq=0$, all these 
configurations are degenerate, lowest-energy, stable solutions. 
Although it may be an interesting exercise to verify their stability 
explicitly by expanding the non-Abelian DBI action to quadratic order in
fluctuations around \eqn{confbis}, one should keep in mind
that this conclusion is only expected to hold at leading order in the
$1/\nc \sim g_s$ expansion, since, as discussed above, we expect 
higher-order effects to lead to an attractive force between the branes,
which would tend to align them.

%%%%%%%%%%%%%%%%%%%%%%%%%%%%%%%%%%%%%%%%%%%%%%%%%%%%%%%%%%%%%%%%%%%%%%
%%%%%%%%%%%%%%%%%%%%%%%%%%%%%%%%%%%%%%%%%%%%%%%%%%%%%%%%%%%%%%%%%%%%%%
%%%%%%%%%%%%%%%%%%%%%%%%%% SECTION %%%%%%%%%%%%%%%%%%%%%%%%%%%%%%%%%%%
%%%%%%%%%%%%%%%%%%%%%%%%%%%%%%%%%%%%%%%%%%%%%%%%%%%%%%%%%%%%%%%%%%%%%%
%%%%%%%%%%%%%%%%%%%%%%%%%%%%%%%%%%%%%%%%%%%%%%%%%%%%%%%%%%%%%%%%%%%%%%
\section{Finite temperature physics} \label{fintem}

The gauge theory we have so far been describing is at zero
temperature. To study the theory at a finite temperature $T$, we
may follow the standard prescription: analytically continue the
time coordinate $t\to t_\mt{E}=it$, periodically identify
$t_\mt{E}$ with period $\delta \te = 1/T$, and impose
anti-periodic boundary conditions on the fermions around the
$\te$-circle. This prescription applies equally well in the gauge
theory and in the dual string description. In the latter, however,
one must consider all possible spacetime solutions with the
appropriate boundary conditions, that is, possible saddle points 
of the Euclidean path integral over supergravity (or rather, string)
configurations \cite{Witten98b}.

In the present case, there are two\footnote{In analogy with
\cite{fairies}, there may be an infinite family of Euclidean
solutions labelled by two integers specifying which cycle of the
$\te\tau$-torus shrinks to zero-size in the bulk. However, given
that the $\te$- and $\tau$-axes are orthogonal, here either the
solutions in eq.~\eqn{d4nut} or \eqn{bhmetric} will dominate the
path integral.} Euclidean supergravity solutions which must be
considered. The first is simply the Euclideanised version of that
appearing in eq.~\eqn{metric},
\beq 
ds^2=\left(\frac{U}{R}\right)^{3/2} \left(d\te^2 +
\sum_{i=1}^3 dx^i dx^i + f(U) d\tau^2\right)+
\left(\frac{R}{U}\right)^{3/2}\frac{dU^2}{f(U)}
+R^{3/2}U^{1/2}d\Omega^2_{\it 4}\,,
\label{d4nut} 
\eeq
which dominates at low temperatures. Recall that 
$f = 1 -\ut^3 / U^3$. The second is the Euclidean black hole 
coming from a nonextremal D4-brane throat
\beq
ds^2=\left(\frac{U}{R}\right)^{3/2}\left(\tilde{f}(U)dt_\mt{E}^2+
\sum_{i=1}^3 dx^i dx^i +d\tau^2\right)+
\left(\frac{R}{U}\right)^{3/2}\frac{dU^2}{\tilde{f}(U)}
+R^{3/2}U^{1/2}d\Omega^2_{\it 4} \,, \label{bhmetric} \eeq
which dominates at high temperatures. In this case, $\tilde{f}=1-
\uh^3 / U^3$. In the Lorentzian-signature solution, the event
horizon is located at $U=\uh$.

Of course, these Euclidean metrics are identical up to
interchanging the role of $\te$ and $\tau$ (and replacing $\ut$
with $\uh$). To match the boundary conditions set by the 
finite-temperature gauge theory, we identify these coordinates with
periods
\be 
\d\te  = \fc{1}{T} \sac \d\tau = \fc{2\pi}{\mkk} \,,
\label{periods}
\ee
in both solutions. Since we wish to avoid conical
singularities in either spacetime at $U=\ut, \uh$, the
boundary conditions fix the metric parameters as
\be 
\ut =\left({4\pi\over3\,\d\tau}\right)^2R^3\sac \uh=
\left({4\pi\over3\,\d\te}\right)^2R^3\,. 
\label{parmu}
\ee

As we commented above, the metric \reef{d4nut} dominates the path
integral at low temperatures, whereas that given by eq.~\reef{bhmetric} 
dominates at high temperatures. One determines the
transition point by comparing the Euclidean action of these two
solutions \cite{Witten98b}. As the metrics describe the same
geometry, it is easy to see that the transition occurs precisely when 
$\d \tau=\d \te$. That is, a phase transition occurs at the critical
temperature
\be 
\ct=\mkk/2\pi\,.
\label{phtemp}
\ee
From the gauge theory viewpoint, this is a confinement/deconfinement 
phase transition. This can be seen by the fact that the temporal 
Wilson loop (around $\te$) vanishes for the low-temperature 
background \reef{d4nut}, whereas it is
nonvanishing for the Euclidean black hole \reef{bhmetric}
\cite{Witten98b}. Alternatively, on the Lorentzian sections, the
Wilson loop as computed from the solution \eqn{metric} exhibits an
area law behaviour, while that computed for the Lorentzian black
hole does not because the string can break in two pieces, each
having an endpoint at the horizon \cite{break}. One might also
observe that, in the low-temperature phase, one has a discrete
spectrum of glueballs \cite{COOT98}, but this gives way to a
continuum of excitations in the high-temperature phase dual to the
black hole background.\footnote{Note that the spectrum would again
be calculated with Lorentzian signature.}
The analogous phase transition in 3+1 dimensions was discussed 
in ref.~\cite{phases}. Finally, note that the phase transition takes 
place at a temperature, \reef{phtemp}, where the gauge theory is
starting to behave five-dimensionally, rather than four-dimensionally.

The preceding discussion refers only to the physics of the
(compactified) five-dimensional gauge theory and is independent of
the presence of fundamental matter fields. In previous sections we
have exhaustively analysed the physics of dynamical quarks in the
confining phase. We can extend the discussion to the deconfining
phase by considering probe D6-branes in the black hole background
\eqn{bhmetric}. In reference \cite{BEEGK03} the study of D7-branes
in an asymptotically \ads{5} black hole background revealed the
existence of two sets of regular embeddings, parametrised by the
quark mass and chiral condensate of the dual gauge theory, and
distinguished by whether or not the D7-brane intersects the
horizon. It was suggested that the gauge theory undergoes a phase
transition, signalled by a discontinuity in $dc/dm_\mt{q}$, when
the D7-brane undergoes its `geometric' transition. Here we
consider similar physics in our four-dimensional gauge theory.

As before, the D6-branes span (Euclidean) time and the spatial
123-directions in common with the D4-brane worldvolume, and lie at
a fixed value of $\tau$. In parallel with the steps described by
eqs.~(\ref{isomer}--\ref{isometric}), we introduce isotropic
coordinates $\{\l, \Omega_{\it 2}, r, \phi\}$ for the Euclidean
black hole \reef{bhmetric}. Furthermore,  we rescale to dimensionless
coordinates
\beq 
\l \to \uh\l\ ,\qquad r\to \uh r\ ,\qquad \r\to \uh\r\ , 
\eeq
in analogy with eq.~\reef{tildes}. Now the embedding equation for
$r(\l)$ becomes
\beq
\frac{d}{d\l}\left[\left(1-\left(\frac{1}{4\r^3}\right)^2\right)\l^2
\frac{\dot{r}}{\sqrt{1+\dot{r}^2}}\right]
=\frac{3}{8}\frac{r\l^2}{\r^8}\sqrt{1+\dot{r}^2}\,.
\label{bhembed}
\eeq
Comparing this result with eq.~\reef{resc}, we see that the sign
of the right-hand side has changed. As we will see below, this
change of sign indicates that the brane now bends towards $U=\uh$:
as expected, the black hole exerts an attractive force on
the D6-brane, in contrast to the repulsive force in the 
\nonsol\ background.

If we look for asymptotically constant embeddings such that
$r(\l)\to r_\infty=L/\uh$ as $\l\to\infty$, we find the same
long-distance behaviour as before:
\beq r\simeq
r_\infty+\frac{c}{\l}\ .\label{asymprof} \eeq
These asymptotic parameters are related to the quark mass and
chiral condensate as before, except for the substitution of $\uh$
for $\ut$:
\be \mq = \fc{\uh r_\infty}{2\pi\ell_s^2}\sac \cc =  -8 \pi^2
\ell_s^2 \t6 \uh^2 \, c \,. \label{magain} \ee
Again, the embedding equation can be solved numerically for
any value of $r_\infty$, with the asymptotic boundary conditions
implicit in \reef{asymprof}. Inspection of the results reveals
that, just as in reference \cite{BEEGK03}, two qualitatively
different types of profile are possible: those for which the probe
brane intersects the horizon, and those for which it does not. As
we vary the boundary condition $r_\infty$, we find that there is
a unique embedding for sufficiently large or small values. In the
case of small $r_\infty$, the unique embedding is such that the
D6-brane falls into the horizon, whereas in the large-$r_\infty$
case it does not, despite, of course, still being attracted by the
black hole. For an intermediate range with
$r_\infty\sim 1$, we find that more than one embedding is
possible. In particular, for a given $r_\infty$, we find both
embeddings which intersect the horizon and those where the
D6-brane smoothly closes off before reaching the horizon.
Furthermore, we also find that there are typically
multiple solutions of at least one of these varieties of
embedding; figure \ref{bhprof} displays some representative
examples. These results imply that in varying from large to
small $r_\infty$, there will be a phase transition in the
behaviour of the quarks. This transition will occur in the
intermediate regime, where, for a fixed temperature and quark mass,
the condensate in the gauge theory can have multiple values. Of
course, the true ground-state condensate will be selected by
minimizing the energy density.

\FIGURE{ \epsfig{file=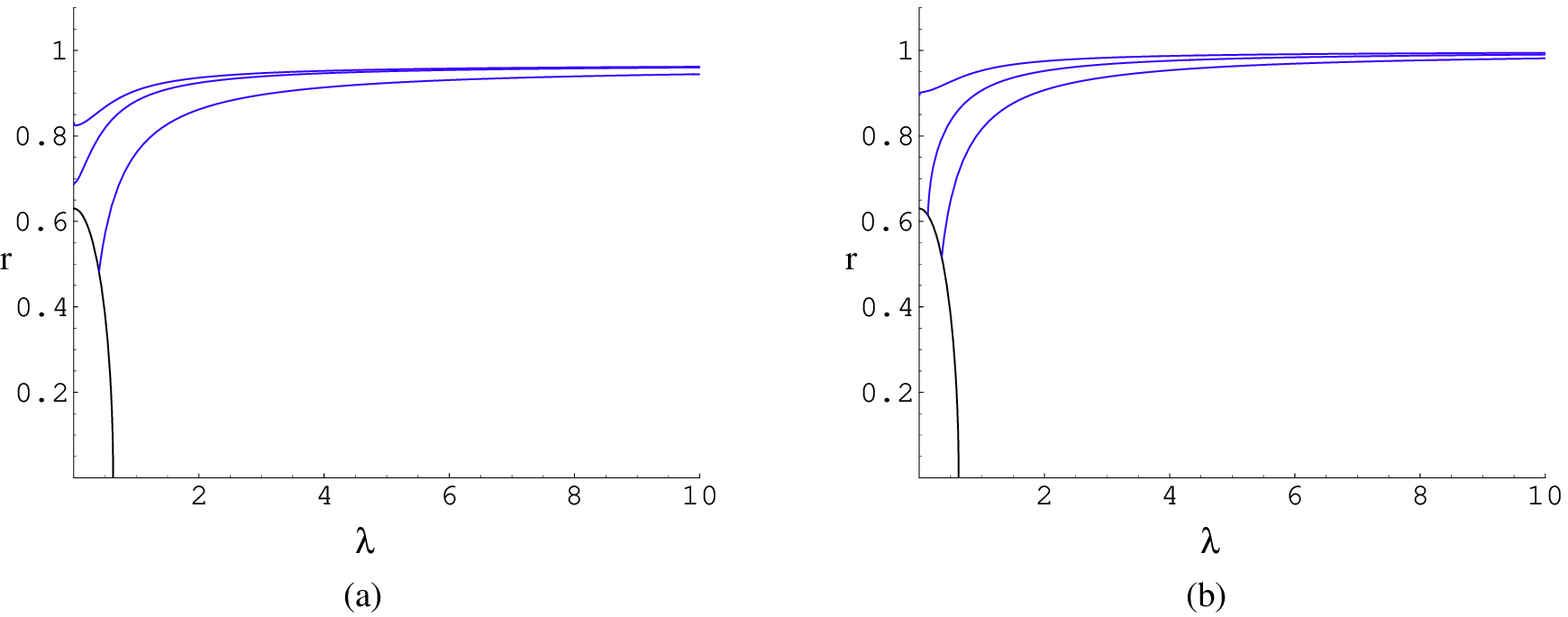, height=6cm} 
\caption{Some representative D6-brane embeddings from the region 
in which $c$ is multi-valued. The three profiles in each figure 
have the same asymptotic value of $r_\infty$; in (a) two of them 
avoid the horizon whereas in (b) two fall into the horizon.}
\label{bhprof} }

Combining various results --- see eqs.~\eqn{inverse}, \eqn{periods}, 
\reef{parmu} and \eqn{magain} --- the relation between the temperature
and $r_\infty$ is found to be
\be 
T = \fc{\tb}{\sqrt{r_\infty}} \sac 
{\rm with}\ \tb^2 \equiv
\frac{9}{4\pi}\frac{m_\mt{q}M_\mt{KK}}{g^2_\mt{YM}N_\mt{c}} \,.
\label{Tfromr} 
\ee
Here $\tb$ can be interpreted, for a given quark mass, as the 
typical meson mass scale in the regime where supersymmetry is 
restored in the zero-temperature gauge theory --- see the 
discussion below eq.~\reef{region2}. In the zero-temperature 
analysis of previous sections we regarded the variation of 
$r_\infty$ as a variation of the quark mass, while $\ut$ or, 
equivalently, the KK scale, was fixed. In this section we 
wish to study the phase structure of the gauge theory as the
temperature is varied, keeping the microscopic gauge theory
parameters, and in particular the quark mass, fixed.
For this reason, since in this phase the physics will only 
depend on the ratio $T/\tb$, as suggested by eq. \eqn{Tfromr}, 
we will think of the variation of $r_\infty$ as a 
variation of the temperature.

\FIGURE{ \epsfig{file=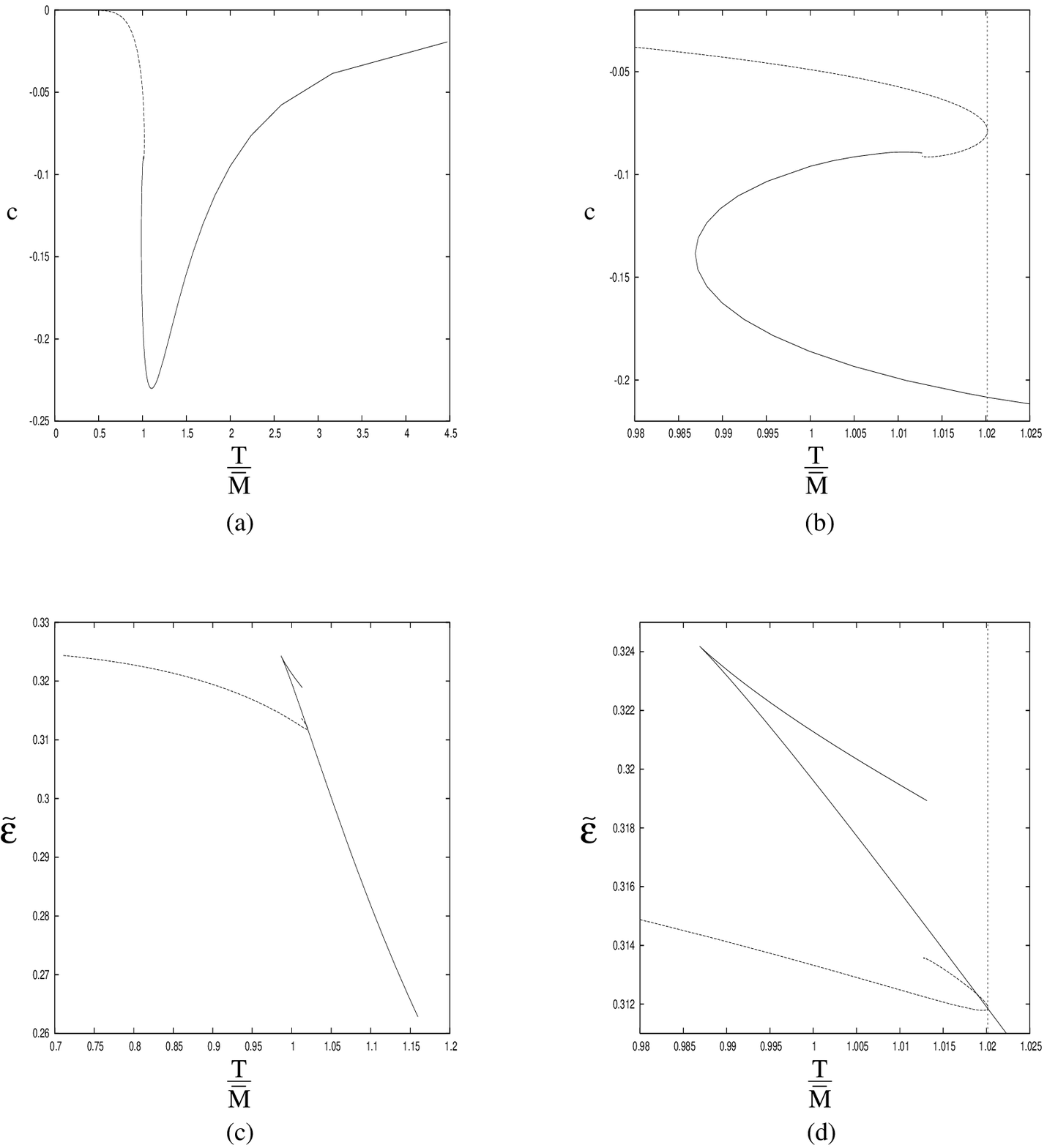, height=15cm} \caption{The quark
condensate $\cc \propto c$ (plots (a) and (b)),
and the energy density $\tilde{\cale}$ (plots (c) and (d)) as
functions of temperature. The solid (dashed) lines describe the
family of embeddings that do (not) intersect the horizon. Plots
(b) and (d) show in more detail the regions of plots (a) and (c)
where the two branches intersect. The vertical line lies at
$T=\tcc$, the temperature at which the phase transition in the
behaviour of the quarks takes place.}
 \label{cvsT} }

The behaviour of the condensate (represented by $c$) as a function
of the temperature is shown in figures \ref{cvsT}a and \ref{cvsT}b. 
These plots
display some interesting features. First note that $c\ra0$ both as
$T\ra0$ and as $T\ra\infty$. In the high-temperature regime, the
unique embedding is such that the D6-brane falls into the
horizon. By following the solid line in the figure, we see
that these embeddings only exist down to some some minimum
temperature, $T\simeq 0.9869 \tb$. Similarly there is a unique
embedding at sufficiently low temperatures such that the D6-brane
does not reach the horizon, and following the dashed line shows
that embeddings of this form only exist up to $T\simeq1.0202\tb$.
As shown most clearly in figure \ref{cvsT}b, multiple solutions
exist in the intermediate temperature range between the two values
given above.

Naturally, where more than one solution to the embedding equation
exists, the ground-state profile will be that selected out by
lowest energy density. Since we are considering static
configurations, the energy density is given by
\beq
\cale=-\int \df\l\df\Omega_{\it 2} {\cal L}= 4\pi
T_\mt{D6}\uh^3 \,\int_{\l_\mt{min}}^\infty\df\l\
\left(1-\left(\frac{1}{4\r^3}\right)^2\right)
\l^2\sqrt{1+\dot{r}^2}\ ,
\label{bhenergy}
\eeq
where $\l_\mt{min}$ is either zero, or the point where the brane
meets the horizon, depending on which variety of embedding we are
considering. This integral is divergent, so to compare the energy
densities of different embeddings we regularize $\cale$ by using as a 
reference the
embedding $r(\l)=0$, which runs from the horizon (at $\l=\l_\mt{H}$, say)
to infinity. Since this is a numerical calculation, we
also introduce a cut-off, $\lm$, in $\l$ at  the point where, by
imposing the appropriate boundary conditions, we match the
numerical profile to the asymptotic solution, $r\simeq r_\infty+c/\l$
. 
There is an
infinite contribution to $\cale$ from the remainder of the profile, but
when regularized the leading behaviour of this correction
can be shown to be
\beq
\cale_\mt{corr}=\frac{c^2}{2\lm}+\calo(\lm^{-3})\ .
\eeq
Therefore, to be precise, we compare the energy densities by calculating
\beqa
\tilde{\cale}=\frac{\cale_\mt{reg}}{4\pi
T_\mt{D6}\uh^3}&\equiv&\int_{\l_\mt{min}}^{\lm} \df\l\
\left(1-\left(\frac{1}{4\r^3}\right)^2\right)\l^2\sqrt{1+\dot{r}^2}\non\\
&& \hspace{1cm} -\int_{\l_\mt{H}}^{\lm} \df\l\
\left(\l^2-\frac{1}{16\l^4}\right)+
\frac{c^2}{2\lm} \,.
\eeqa
We then trust that this provides a valid basis for the comparison of
different energy densities when the differences in $\tilde{\cale}$ are
greater than $\calo(\lm^{-3})$. This is the case here, as we use
$\lm=100$.

The energy density $\tilde{\cale}$ is plotted as a function of temperature
in figures \ref{cvsT}c and \ref{cvsT}d. Starting from low temperatures, figure
\ref{cvsT}d shows that, as $T$ is increased to that in the
intermediate region, the physical embedding remains that which
does not reach the horizon. At $T\simeq 1.0202 \tb$, the energy
density for this embedding and that which falls into the horizon
match. For $T > 1.0202 \tb$, the energetically favoured embedding is
one which falls into the horizon. Therefore, at
\be 
T\equiv\tcc\simeq 1.0202\tb\,, 
\ee
there is a phase transition in the behaviour of the fundamental
quarks. This phase transition is signalled by a discontinuity in
$\langle \bar{\psi} \psi \rangle(T)$, as shown in Figure
\ref{cvsT}b, as well as a discontinuity in the specific heat
$d\tilde{\cale}(T)/dT$, as can be seen in Figure \ref{cvsT}d. Note
that, to within the precision of our numerical calculations, $\tcc$ 
cannot be distinguished from the maximum temperature for which 
a horizon-avoiding embedding is possible.  

From the string description with the D4 throat geometry, the basic
physics behind this transition is straightforward. Increasing the
temperature increases both the radial position of the horizon and
energy density of the black hole. For  small enough a
temperature, the D6-brane is pulled towards the horizon but
its tension balances the gravitational attraction. However, above
the critical temperature $\tcc$, the gravitational force has
increased to the point where it overcomes the tension and the
brane is drawn into the horizon.

Another feature that characterizes the present phase transition is
that the meson spectrum changes at $T=\tcc$. The discrete spectrum of
mesons of the low-temperature phase becomes a continuum 
of excitations in the high-temperature phase where the D6-brane 
embedding extends to the horizon. This behaviour is similar to that of 
the glueball spectrum at the deconfinement phase transition in the 
gauge theory, discussed previously. 
Note, however, that for temperatures above $\ct$ we are, by 
assumption, in the deconfined phase of the gauge theory, and so it is 
possible to introduce free quarks; these correspond to open strings 
extending from the D6-brane down to the horizon. If $\ct < \tcc$ there 
is then a range of temperatures, $\ct < T < \tcc$, in which there are
free quarks but also a finite mass gap, $m_\mt{eff}(T)>0$, 
associated with the introduction of such quarks into the 
system. In this regime,
there still exists a (discrete) spectrum of bound states with 
$M<2m_\mt{eff}$. Note also that the deconfined adjoint degrees of
freedom still generate a finite screening length beyond which the force 
between a quark and an anti-quark vanishes \cite{break}; in the 
string description this happens because the string that joins the 
quarks together snaps into two pieces, each of them going from the
brane to the horizon, that can be separated with no cost in energy. 
Above $\tcc$, however, $m_\mt{eff}(T)=0$ and the spectrum of
fundamental matter field excitations becomes continuous. 

Up to this point, we have discussed two independent phase
transitions: the deconfining phase transition at $T=\ct$, for
which the holographic dual description involved a discrete change
of the bulk spacetime geometry, and the quark phase transition at
$T=\tcc$, for which the holographic description involved a
discrete jump in the embedding geometry of the D6-brane. However,
if the latter is to be realized as a separate phase transition,
then it must be true that $\tcc>\ct$.\footnote{Note that, in this
regime, the fundamental fields still `feel' the deconfining
transition at $T=\ct$. For example, there is a discontinuity in
the chiral condensate $\cc$, which jumps from a negative to a
positive value at this phase transition.} 
In terms of the microscopic parameters, this condition translates into
\beq 
m_\mt{q} > {0.98\over 9\pi}g^2_\mt{YM} N_\mt{c} M_\mt{KK}\ . 
\label{phtemp2}
\eeq
In the zero-temperature theory, this is roughly the quark mass
scale at which supersymmetry is restored --- see the discussion below
eq.~\reef{region2}.

If $\tcc<\ct$, the relevant spacetime geometry at $T=\tcc$ is the
\nonsol{} \reef{d4nut} and the ambiguity in the D6-brane embedding
discussed above is not relevant. Alternatively, the gauge theory
is still in a confined phase and so the chemical potential to
introduce a free quark remains infinite; in the dual description this
is because, since there is no horizon, an open string must have both
ends on the D6-brane, and so represents a quark/anti-quark bound
state. In this case, the two phase transitions are realized
simultaneously at $T=\ct$. An important example of this behaviour is 
when $\mq=0$. In this case, the low-temperature phase is characterized 
by spontaneous chiral symmetry breaking, as in section \ref{broken}.
In the high-temperature phase, $\mq \ra 0$ is equivalent to 
$T \ra \infty$ (see eq. \eqn{Tfromr}), in which limit the chiral
condensate vanishes (see figure \ref{cvsT}a). Hence the deconfining 
phase transition is also accompanied by the restoration of chiral 
symmetry in this model.

To close the section, we wish to stress to the reader that while
these phase transitions are of inherent interest in understanding the
model under study here, they take place in a temperature regime
where the gauge theory is intrinsically five-dimensional, rather
than behaving in a four-dimensional way, as can be seen by
comparing the transition temperatures given in eqs.~\reef{phtemp}
and \reef{phtemp2} to the compactification scale $\mkk$.

%%%%%%%%%%%%%%%%%%%%%%%%%%%%%%%%%%%%%%%%%%%%%%%%%%%%%%%%%%%%%%%%%%%%%%%%%%%%%%
%%%%%%%%%%%%%%%%%%%%%%%%%%%%%%%%%%%%%%%%%%%%%%%%%%%%%%%%%%%%%%%%%%%%%%%%%%%%%%
%%%%%%%%%%%%%%%%%%%%%% SECTION  %%%%%%%%%%%%%%%%%%%%%%%%%%%%%%%%%%%%%%%%%%%%%
%%%%%%%%%%%%%%%%%%%%%%%%%%%%%%%%%%%%%%%%%%%%%%%%%%%%%%%%%%%%%%%%%%%%%%%%%%%%%%
%%%%%%%%%%%%%%%%%%%%%%%%%%%%%%%%%%%%%%%%%%%%%%%%%%%%%%%%%%%%%%%%%%%%%%%%%%%%%%
\section{D4/D6/$\overline{\mbox{D6}}$-intersections}
\label{Dbar}

In the discussion of embeddings in section \ref{broken}, we noted
that the solution $\rt (\lt) =0$ of eq.~\eqn{resc} is not a
physical one (for $\ut \neq 0$) if it terminates at the origin of the
$U\tau$-plane because it would correspond to an open D6-brane.
However, this solution can be extended through this point to
construct a physically acceptable embedding. To see this, recall
that the full D6-brane embedding is further specified by the
conditions $\phi=\phi_0$ and $\tau=\tau_0$, where $\phi_0$ and
$\tau_0$ are arbitrary constants. We can therefore join together a
solution with $\tau=\tau_0$ to another one with $\tau=\tau_0+ \d
\tau/2$ (see eq.~\eqn{deltatau}) to obtain a complete regular
solution that describes a D6-brane whose projection on the
$U\tau$-plane is a straight line that passes through the origin
and intersects the boundary at two antipodal values of $\tau$.
This means that this solution describes an intersection of $\nc$
D4-branes with two D6-branes, or, more precisely, with one
D6-brane and one anti-D6-brane.\footnote{Technically, the reason
is that the orientations on the intersections of the D6-brane with
the boundary at $\tau=\tau_0$ and $\tau=\tau_0+ \d \tau/2$,
induced from a given orientation on the D6-brane, are opposite to
each other.} Note that these embeddings bear a striking resemblance to
those discussed in \cite{SS03}. The parallel includes the facts that
both constructions involve two boundary defects and both preserve
the chiral symmetry.

To better understand the defect theory corresponding to these
D4/D6/$\overline{\mbox{D6}}$-intersections, we consider a family
of more general embeddings that describe D6-branes that lie at
$\rt=0$ (or alternatively, $z^8=0=z^9$) and stretch along some
curve $U=U(\tau)$ in the $U\tau$-plane, intersecting the boundary
at two, not necessarily antipodal, values of $\tau$. Such
D6-branes wrap a maximal $S^2$ within the $S^4$ in the metric
\eqn{metric}, so the induced metric on their worldvolume is
\be 
ds^2(g) = \left( \fc{U}{R} \right)^{3/2} \, 
\eta_{\mu\nu} dx^\mu dx^\nu +
R^{3/2} U^{1/2} \, d\Omega_{\it 2}^2 + \left[ \left( \fc{U}{R}
\right)^{3/2} \, f(U) + \fc{U'^2}{\left( \fc{U}{R} \right)^{3/2}
\, f(U)} \right] \, d\tau^2 \,. 
\ee
Since the Lagrangian density on the D6-brane
\be 
\call_{\mt{D6}} = - \fc{1}{(2\pi)^6 \ell_s^7} \, 
e^{-\phi} \, \sqrt{-\det g_{\mt{ind}}} = 
\t6 \, U^2 \, \sqrt{\left( \fc{U}{R} \right)^3 f(U) + \fc{U'^2}{f(U)}} 
\ee
(where $U' = dU/d\tau$) does not depend explicitly on $\tau$, 
it follows that
\be 
U' \, \fc{\pa \call}{\pa U'} - \call = \fc{U^2 \left( \fc{U}{R}
\right)^3 f(U)} {\sqrt{\left( \fc{U}{R} \right)^3 f(U) +
\fc{U'^2}{f(U)}}} 
\ee
is also $\tau$-independent. Let $U_0 \geq \ut$ be the minimum
value of $U(\tau)$. By symmetry we can assume that this value is
reached at $\tau=0$, that is, that we have $U(0)=U_0$ and
$U'(0)=0$. Integrating the equation above we then find
\be
\tilde{\tau} (u) = \fc{3}{2} \, b^{7/2} \, g^{1/2}(b) \, \int_b^u
dx \fc{1}{x^{3/2} g(x) \sqrt{x^7 g(x) - b^7 g(b)}} \,, \label{tau}
\ee
where we have introduced
\be 
\tilde{\tau} \equiv \fc{3}{2} \,
\fc{\ut^{1/2}}{R^{3/2}} \, \tau \sac u \equiv \fc{U}{\ut} \sac b
\equiv \fc{U_0}{\ut} \geq 1 \sac g(x) \equiv 1 - \fc{1}{x^3} \,.
\ee
Note that the rescaled angle $\tilde{\tau}$ has period $2\pi$.
Eq.~\eqn{tau} specifies the D6-brane profile in the $U\tau$-plane.
The D6-brane intersects the boundary at $\tilde{\tau}_\pm (b) =
\pm \tilde{\tau}(u \ra \infty)$. These two points are implicitly
fixed by $U_0$, the `lowest' point of the D6-brane in the
$U\tau$-plane. In particular, their separation is a
monotonically decreasing function of $b$. It is easy to show that
$\tilde{\tau}_\pm (b \ra 1) = \pm \pi/2$, so the two points
become antipodal as the lowest point of the D6-brane approaches
the origin of the $U\tau$-plane. In this limit the projection of
the D6-brane on this plane is just a straight line, as discussed
above. In the opposite limit in which the (lowest point of)
D6-brane approaches the boundary, the separation between the
intersection points becomes smaller and smaller. In the strict
limit $b\ra \infty$ the two coincide and the D6-brane disappears.

Now, these embeddings seem to portend fascinating physics for the
dual five-dimensional gauge theory coupled to two defects or, more
precisely, to a defect and an anti-defect. It appears, however, that 
this physics is intrinsically five-dimensional. 
Since the focus of the present
paper is the four-dimensional QCD-like physics of the original
model, we have not fully explored this system. However, we offer
some preliminary remarks in the following.

The deflection of the D6-brane in the $\tau$-direction, which
corresponds to a worldvolume direction of the D4-branes, is
reminiscent of certain D5-brane embeddings in \ads{5} studied in
ref.~\cite{ST02}. These authors found (non-supersymmetric)
embeddings in which the D5-brane began at large radius running
along $r$, but were deflected to run parallel to the \ads{5}
horizon at some finite radius. At large $r$, these configurations can
be understood as D3/D5 intersections that are T-dual to that
represented in \eqn{intersection}. The dual gauge theory involved
coupling $\caln=4$ SYM to a ($2+1$)-dimensional defect which broke 
all supersymmetries. The deflection was interpreted as giving an 
expectation value to an operator in the defect theory. 

In the present case, the \nonsol\ background breaks the
supersymmetry but with both defects, supersymmetry would still not
be restored in the limit $\ut\ra0$;\footnote{Note that the
supersymmetry breaking of the background is not essential to
constructing the general embeddings, \ie, one can simply replace
$g(x)$ by 1 everywhere in eq.~\reef{tau}.} 
one way to understand this in the field theory is to note that the half 
of the supersymmetries of the ambient theory preserved (broken) 
by the defect are precisely those that are broken (preserved) by the
anti-defect.\footnote{Defect/anti-defect configurations, and more
  general multiple defect configurations, were discussed in \cite{MNT02}.} 
The embeddings described by eq.~\reef{tau} would correspond
to a state where a symmetric pair of operators in each defect
theory simultaneously acquire identical expectation values.
Indirectly, the D6-brane embedding would also induce different
expectation values of gauge theory operators, \eg, Tr $F^2$, in the
two regions on the $\tau$-circle divided by the defects.

At the microscopic level, there are two sets of fundamental matter 
fields in the field theory, one associated with each defect.
This is realised in the string description by the fact that,
despite there being a single D6-brane, its intersection with a
constant-$U > U_0$ slice consists of two disconnected pieces, 
corresponding to two independent sets of active degrees of freedom 
at the corresponding energy scale. These degrees of freedom seem to
disappear for energy scales associated to radial positions $U < U_0$, 
as in \cite{ST02}. As discussed above, the D6-brane disappears in the limit 
$\tilde{\tau}_+-\tilde{\tau}_- \ra 0$. Consistently, in this limit 
$U_0 \ra \infty$ and the degrees of freedom above are completely
removed. In the gauge theory, this corresponds to the limit in which
the defect and the anti-defect are brought on top of each other.

This situation may be compared with that where both
defects are associated with a `standard' D6-brane embedding as
described in section \ref{broken}; one at 
$\tilde{\tau}= \tilde{\tau}_+$  and the other at 
$\tilde{\tau}= \tilde{\tau}_-$. Of course, in this case there are 
also two sets of independent degrees of freedom associated with open 
string excitations on both the D6- and 
$\overline{\mbox{D6}}$-brane.\footnote{Note that in this case one 
may also realise a defect/defect (as opposed to a defect/anti-defect) 
theory by considering two D6-branes.} However, one difference is that
the spectrum of these excitations will now be independent of the 
defect/anti-defect separation,
$\tilde{\tau}_+-\tilde{\tau}_-$.\footnote{In the approximation in
  which interactions between the branes, which are suppressed in the
  large-$\nc$ limit, are neglected.}  
In addition, one would also find excitations corresponding to open 
strings stretching between the two separate branes. 
In contrast, in the case of a single D6-brane, the two defects can 
interact through open string modes that propagate along the brane from 
one defect to the other. 

Of course, one should ask the question which embedding minimizes
the energy density of the system. That is, does energetics favour
the smooth embedding which joins the two boundary defects or that
in which there are two independent branes? As a preliminary step
in this direction, we compared the energy density for the
`standard' embeddings of the D6- and $\overline{\mbox{D6}}$-brane
pair with $r_\infty=0$ to that of the smooth embedding above with
$U_0=\ut$ (and hence $\tilde{\tau}_+-\tilde{\tau}_- =\pi$). Our
numerical results indicate that the smooth embedding produces a
lower energy density and hence describes the true vacuum
configuration. Note then that with both defects corresponding to
the D6- and $\overline{\mbox{D6}}$-branes, the vacuum preserves
chiral symmetry.

Presumably if $U_0$ is increased above $\ut$, the smooth
embeddings described above would continue to be energetically
favored as it seems the energy density for these solutions must
decrease as $\tilde{\tau}_+-\tilde{\tau}_-$ decreases while that
for the standard embeddings remains unchanged. Note, however, that
the above are just two particular cases of a more general family
of embeddings in which the brane bends simultaneously in the 89-
and the $U\tau$-planes, that is, embeddings with nontrivial
functions $r=r(\l)$ and $\tau=\tau(\l)$ (using the notation of
section 2). In general, we expect the minimum-energy two-defect
embedding is of this type.

These defect theories might be explored in other ways as well. For
example, one might also consider varying the parameter $\r_\infty$
for the two defects, which may be done independently for each of
the defects. Finally, of course, one might also consider the
effects of a finite temperature, as in section 6.

%%%%%%%%%%%%%%%%%%%%%%%%%%%%%%%%%%%%%%%%%%%%%%%%%%%%%%%%%%%%%%%%%%%%%%%%%%
%%%%%%%%%%%%%%%%%%%%%%%%%%%%%%%%%%%%%%%%%%%%%%%%%%%%%%%%%%%%%%%%%%%%%%%%%%
%%%%%%%%%%%%%%%%%%%%%% SECTION  %%%%%%%%%%%%%%%%%%%%%%%%%%%%%%%%%%%%%%%%%%
%%%%%%%%%%%%%%%%%%%%%%%%%%%%%%%%%%%%%%%%%%%%%%%%%%%%%%%%%%%%%%%%%%%%%%%%%%
%%%%%%%%%%%%%%%%%%%%%%%%%%%%%%%%%%%%%%%%%%%%%%%%%%%%%%%%%%%%%%%%%%%%%%%%%%
\section{Discussion} \label{discus}

We have analyzed in detail the case of a single D6-brane, which
corresponds to a single flavour in the gauge theory.
In this case the gauge theory enjoys (at large $\nc$) an exact
$U(1)_\mt{V} \times \ua$ symmetry. The first factor is just the flavour
symmetry, which in the string description corresponds to the gauge
symmetry on the worldvolume of the D6-brane. The second factor is
a chiral symmetry that rotates the left- and right-handed
fermions, $\psi_\mt{L}$ and $\psi_\mt{R}$, with opposite phases;
it also rotates by a phase the adjoint scalar $X=X^8 + i X^9$,
as is clear from the fact that in the string description the
$\ua$ symmetry is just the rotational symmetry in the 89-plane.

Like in QCD, we expect the $\ua$ symmetry to be spontaneously
broken by a chiral condensate, $\cc \neq 0$, and hence the
existence of the corresponding Goldstone boson in the spectrum;
this is the analog of the QCD $\eta'$ particle (which is massless in
the $\nc \ra \infty$ limit), but, in an abuse of language, we have
referred to it as a `pion'. The string description indeed confirms
the field theory expectation: for zero quark mass, there is precisely
one massless pseudo-scalar (in the four-dimensional sense) fluctuation of the
D6-brane. If $\mq> 0$ this pion becomes a pseudo-Goldstone boson. Its
mass, as well as the chiral condensate and the pion decay constant,
can be computed in the string description, and we have shown that
these three quantities and the quark mass obey the
Gell-Mann--Oakes--Renner relation \eqn{GMOR} \cite{GMOR68}.
Although this is a reassuring result, one should keep in mind that,
just like in field theory, it is essentially a consequence of the
{\it symmetry}-breaking pattern.

In the case $\nf >1$ we provided a holographic version of the
Vafa--Witten theorem, which states that, in a vector-like theory
with zero $\theta$-angle (like QCD) and $\nf$ quark flavours of
identical masses $\mq >0$, the global, $U(\nf)$ flavour
symmetry cannot be spontaneously broken. As we discussed,
it is a priori unclear, on field theory grounds, that the Vafa--Witten
theorem does apply to the actual gauge theory on the D4-branes, because
of the presence of Yukawa-like couplings to scalar fields
of masses $\mkk \sim \Lqcd$. The result from the string analysis
therefore established that the presence of these fields does not alter
the vacuum structure of the theory to the extent of violating the
Vafa--Witten theorem.

The global $U(\nf)$ symmetry of the gauge theory becomes the gauge
symmetry on the worldvolume of $\nf$ D6-branes in the string
description. The holographic version
of the Vafa--Witten theorem is the statement that the minimum-energy
configuration for the $\nf$ D6-branes is a $U(\nf)$-preserving one in
which all branes lie on top of each other. It should be emphasised
that, unlike the Gell-Mann--Oakes--Renner relation, the Vafa--Witten
theorem is not based on a symmetry(-breaking) argument, but depends on the
{\it dynamics} of the theory. In the field theory, the proof relies
on the reality and positivity of the fermion determinant that arises
in the path integral formulation; in the holographic version, it
relies on details of the dynamics of multiple D6-branes.

One result of our analysis of the $\nf>1$ case is the existence of
$\nf^2$ independent, massless fluctuations of the D6-branes if $\mq=0$.
It is tempting to interpret the associated $\nf^2$ massless pseudo-scalar
particles in the spectrum of the gauge theory as the Goldstone bosons
of a spontaneously broken $U(\nf)_\mt{A}$ symmetry, to which the
$\ua$ symmetry would putatively be enhanced for $\nf >1$. However,
only one of the $\nf^2$ states above is a true Goldstone boson,
for two reasons. First, the five-dimensional theory contains a
term of the form $\bar\psi_i \, X \, \psi^i$, where $i=1, \ldots, \nf$
is the flavour index. Since $X$ only transforms under the $\ua$
subgroup of $U(\nf)_\mt{A}$, the theory is only invariant under
$\ua \times U(\nf)_\mt{V}$, which is spontaneously broken to
$U(\nf)_\mt{V}$. Although $X$ is a massive field that can
be integrated out, its mass $M_X \sim M_\mt{KK} \sim \Lambda\mt{QCD}$
is of the same order as the strong coupling scale, so its integration
should not lead to a real suppression of the symmetry-breaking operator
above.

The second reason that makes implausible the interpretation of all
but one of the massless modes above as Goldstone bosons is that
they have non-derivative interactions. Indeed, the effective,
interacting Lagrangian for these modes is obtained by expanding
the non-Abelian Dirac--Born--Infeld action for the $\nf$ D6-branes
\cite{Myers99} beyond quadratic order and then dimensionally
reducing it to four dimensions, as we did in Section 4 in the
$\nf=1$ case. Since the action for the D6-branes contains
non-derivative interactions, such as commutator terms squared, so
does the effective four-dimensional Lagrangian. These terms are
only absent for the mode that multiplies the identity generator of
$U(\nf)$. The caveat as to why this argument makes the Goldstone boson
interpretation for the other modes implausible, but not strictly
impossible, is that the coefficients in the effective Lagrangian
might conspire so that, once all contributing diagrams are
included, only momentum-dependent parts survive in any scattering
amplitude (as in the linear sigma-model for the pion Lagrangian
\cite{DGH92}). This possibility seems rather unlikely.

Thus we conclude that the string description predicts the existence
of $\nf^2$ massless pseudo-scalar particles in the spectrum of the
quantum gauge theory in the large-$\nc$ limit. Once sub-leading
effects in the $1/\nc$ expansion are included, we expect these
particles to acquire a small mass. The $\ua$ symmetry is anomalous at
finite $\nc$, so its associated Goldstone boson is not exactly
massless at finite $\nc$; presumably, this anomaly can be analysed in
the supergravity description along the lines of 
\cite{BDFLM01,KOW02,BDFLM02,GHP03}.
The other $\nf^2 -1$ modes are likely to acquire a mass due to closed
string interactions between the D6-branes, which are a
$g_s \sim 1/\nc$ effect. From the gauge theory viewpoint, we have found
no obvious reason for the masslessness (or lightness, at finite $\nc$)
of these $\nf^2 -1$ pseudo-scalars, since we have argued that they are not
Goldstone bosons.\footnote{Other examples with `unexpectedly' light scalars in
the context of AdS/CFT have been discussed in \cite{Strassler03}.}
Recall that, from the string viewpoint, the reason for this is that
the only difference between the non-Abelian and the Abelian DBI
actions consists exclusively of terms that involve commutators of
fields. This can be regarded as a consequence of the fact that the
seven-dimensional D6-brane action is T-dual to the ten-dimensional
action of D9-branes; all commutators of scalar fields in the D6-brane
action originate from commutators of gauge fields in the non-Abelian field
strength on the D9-branes. Thus, it is {\it ten}-dimensional $U(\nf)$
gauge invariance that seems to be responsible, from the string
viewpoint, for the lightness of the gauge theory scalars.\footnote{
This observation is due to M.\ Strassler.} Given the
importance in high energy physics of mechanisms that naturally allow
small scalar masses, it would be interesting to understand how to
rephrase the constraints imposed by ten-dimensional $U(\nf)$
gauge invariance purely in four-dimensional field theory terms.

%%%%%%%%%%%%%%%%%%%%%%%%%%%%%%%%%%%%%%%%%%%%%%%%%%%%%%%%%%%%%%%%%%%
%%%%%%%%%%%%%%%%%%%%%%%%%%%%%%%%%%%%%%%%%%%%%%%%%%%%%%%%%%%%%%%%%%%
%%%%%%%%%%%%%%%%%%%%  SECTION   %%%%%%%%%%%%%%%%%%%%%%%%%%%%%%%%%%
%%%%%%%%%%%%%%%%%%%%%%%%%%%%%%%%%%%%%%%%%%%%%%%%%%%%%%%%%%%%%%%%%%%
%%%%%%%%%%%%%%%%%%%%%%%%%%%%%%%%%%%%%%%%%%%%%%%%%%%%%%%%%%%%%%%%%%%

%\section*{Acknowledgements}
\acknowledgments

We are especially grateful to J.\ Brodie for helpful
conversations. We also wish to thank J.\ Donoghue, J.\ Erdmenger,
J.\ I.\ Latorre, A.\ E.\ Nelson, D.\ T.\ Son, M.\ J.\ Strassler,
S.\ Thomas, D.\ Tong and A.\ Uranga for discussions and comments. 
Research at
the Perimeter Institute is supported in part by funds from NSERC
of Canada. RCM is further supported by an NSERC Discovery grant,
DJW by Fonds FCAR du Qu\'ebec and by a McGill
Major Fellowship, while MK is supported in part by NSF grants
PHY-0331516, PHY99-73935 and DOE grant DE-FG02-92ER40706. 
DJW wishes to thank the University of Waterloo Physics
Department for their ongoing hospitality and MK would like to thank
the Perimeter Institute for Theoretical Physics for partial
support and hospitality while this work was being performed.


\begin{thebibliography}{99}

\bibitem{Maldacena97}
J.\ M.\ Maldacena, {\it The large N limit of superconformal field
theories and supergravity}, \atmp{2}{1998}{231}
[\ijtp{38}{1998}{1113}], \hepth{9711200}.

\bibitem{Witten98b}
E.\ Witten, {\it Anti-de Sitter space, thermal phase transition,
and confinement in gauge theories}, \atmp{2}{1998}{505},
\hepth{9803131}.

\bibitem{COOT98}
C.\ Csaki, H.\ Oouguri, Y.\ Oz and J.\ Terning,
{\it Glueball mass spectrum from supergravity},
\jhep{01}{1999}{017}, \hepth{9806021};
R.\ de Mello Koch, A.\ Jevicki, M.\ Mihailescu and J.\ P.\ Nunes,
{\it Evaluation of glueball masses from supergravity},
\prd{58}{1998}{105009}, \hepth{9806125};
H.\ Ooguri, H.\ Robins and J.\ Tannenhauser,
{\it Glueballs and their Kaluza-Klein cousins},
\plb{437}{1998}{77}, \hepth{9806171};
J.\ A.\ Minahan,
{\it Glueball mass spectra and other issues for supergravity duals
of QCD models}, \jhep{01}{1999}{020}, \hepth{9811156};
N.\ R.\ Constable and R.\ C.\ Myers,
{\it Spin-two glueballs, positive energy theorems and
the AdS/CFT correspondence}, \jhep{10}{1999}{037},
\hepth{9908175};
R.\ C.\ Brower, S.\ D.\ Mathur and C.\ Tan,
{\it Glueball spectrum for QCD from AdS supergravity duality},
\npb{587}{2000}{249}, \hepth{0003115}.

\bibitem{AFM98}
O.\ Aharony, A.\ Fayyazuddin and J.\ M.\ Maldacena,
{\it The large-N limit of ${\cal N} =2,1$ field theories
from threebranes in F-theory}, \jhep{07}{1998}{013},
\hepth{9806159}.

\bibitem{KR01}
A.\ Karch and L.\ Randall,
{\it Open and closed string interpretation of SUSY CFT's on
branes with boundaries}, \jhep{06}{2001}{063}, \hepth{0105132}.

\bibitem{BDFLM01}
M.\ Bertolini, P.\ Di Vecchia, M.\ Frau, A.\ Lerda, and R.\ Marotta,
{\it N=2 Gauge theories on systems of fractional D3/D7 branes},
\npb{621}{2002}{157}, \hepth{0107057}.    

\bibitem{BDFM01}
M.\ Bertolini, P.\ Di Vecchia, G.\ Ferretti, and R.\ Marotta,
{\it Fractional Branes and N=1 Gauge Theories},
\npb{630}{2002}{222}, \hepth{0112187}.

\bibitem{KK02}
A.\ Karch and E.\ Katz, {\it Adding flavor to AdS/CFT},
\jhep{06}{2002}{043}, \hepth{0205236}.

\bibitem{KKW02}
A.~Karch, E.~Katz and N.~Weiner, {\it Hadron masses and screening
from AdS Wilson loops}, \prl{90}{2003}{091601}, \hepth{0211107}.

\bibitem{WH03}
X.~J.~Wang and S.~Hu, {\it Intersecting branes and adding flavors
to the Maldacena-Nunez background}, \jhep{09}{2003}{017},
\hepth{0307218}.

\bibitem{Ouyang03}
P.~Ouyang, {\it Holomorphic D7-branes and flavored N = 1 gauge
theories}, \hepth{0311084}.

\bibitem{KMMW03}
M.\ Kruczenski, D.\ Mateos, R.\ C.\ Myers and D.\ J.\ Winters,
{\it Meson spectroscopy in AdS/CFT with flavour},
\jhep{07}{2003}{049}, \hepth{0304032}.

\bibitem{SS03}
T.\ Sakai and J.\ Sonnenschein,
{\it Probing flavored mesons of confining gauge theories by
supergravity}, \jhep{09}{2003}{047}, \hepth{0305049}.

\bibitem{BEEGK03}
J.\ Babington, J.\ Erdmenger, N.\ Evans, Z.\ Guralnik and I.\ Kirsch,
{\it Chiral symmetry breaking and pions in non-supersymmetric
gauge/gravity duals}, \hepth{0306018}.

\bibitem{NPR03}
C.\ Nunez, A.\ Paredes and A.\ V.\ Ramallo,
{\it Flavoring the gravity dual of $\caln=1$ Yang-Mills with probes},
\jhep{12}{2003}{024}, \hepth{0311201}.

\bibitem{IMSY98}
N.\ Itzhaki, J.\ M.\ Maldacena, J.\ Sonnenschein and S.\ Yankielowicz,
{\it Supergravity and the large N limit of theories with sixteen
supercharges}, \prd{58}{1998}{046004}, \hepth{9802042}.

\bibitem{DFO01}
O.\ DeWolfe, D.\ Z.\ Freedman and H.\ Ooguri,
{\it Holography and Defect Conformal Field Theories},
\prd{66}{2002}{025009}, \hepth{0111135}.

\bibitem{GMOR68}
M.\ Gell-Mann, R.\ J.\ Oakes and B.\ Renner,
{\it Behavior of current divergences under $SU(3) \times SU(3)$},
\pr{175}{1968}{2195}.

\bibitem{VW84}
C.\ Vafa and E.\ Witten, {\it Restrictions on symmetry breaking
in vector-like gauge theories}, \npb{234}{1984}{173}.

\bibitem{SVZ79}
M.\ A.\ Shifman, A.\ I.\ Vainshtein and V.\ I.\ Zakharov,
{\it QCD and resonance physics. Sum rules.},
\npb{147}{1979}{385}.

\bibitem{soliton}
G.~T.~Horowitz and R.~C.~Myers, {\it The AdS/CFT correspondence
and a new positive energy conjecture for  general relativity},
Phys.\ Rev.\ {\bf D 59} (1999) 026005, \hepth{9808079}.

\bibitem{adsprop}
U.~H.~Danielsson, E.~Keski-Vakkuri and M.~Kruczenski, {\it Vacua,
propagators, and holographic probes in AdS/CFT}, JHEP {\bf 9901}
(1999) 002, \hepth{9812007}.

\bibitem{Myers99}
R.\ C.\ Myers, {\it Dielectric-Branes}, \jhep{12}{1999}{022},
\hepth{9910053}.

\bibitem{fairies}
R.\ Dijkgraaf, J.\ M.\ Maldacena, G.\ W.\ Moore and E.\ Verlinde,
{\it A black hole farey tail}, \hepth{0005003}.

\bibitem{phases}
R.\ C.\ Myers, {\it Stress tensors and Casimir energies
in the AdS/CFT correspondence}, \prd{60}{1999}{046002},
\hepth{9903203}; 
S.\ Surya, K.\ Schleich and D.\ M.\ Witt,
{\it Phase transitions for flat adS black holes},
\prl{86}{2001}{5231}, \hepth{0101134}.

\bibitem{break}
S.\ J.\ Rey, S.\ Theisen and J.\ T.\ Yee,
{\it Wilson-Polyakov loop at finite temperature in large 
N gauge theory and anti-de Sitter supergravity}, 
\npb{527}{1998}{171}, \hepth{9803135}.

\bibitem{ST02}
K.\ Skenderis and M.\ Taylor,
{\it Branes in AdS and pp-wave spacetimes},
\jhep{06}{2002}{025}, \hepth{0204054}.

\bibitem{MNT02}
D.\ Mateos, S.\ Ng and P.\ K.\ Townsend,
{\it Supersymmetric Defect Expansion in CFT from AdS Supertubes},
\jhep{07}{2002}{048}, \hepth{0207136}.

\bibitem{DGH92}
J.\ F.\ Donoghue, E.\ Golowich and B.\ R.\ Holstein,
{\it Dynamics of the standard model},
Cambridge University Press, 1992.

\bibitem{KOW02}
I.\ R.\ Klebanov, P.\ Ouyang and E.\ Witten,
{\it A gravity dual of the chiral anomaly},
\prd{65}{2002}{105007}, \hepth{0202056}.

\bibitem{BDFLM02}
M.\ Bertolini, P.\ Di Vecchia, M.\ Frau, A.\ Lerda and R.\ Marotta,
{\it More Anomalies from Fractional Branes}, \plb{540}{2002}{104},
\hepth{0202195}.

\bibitem{GHP03}
U.\ Gursoy, S.\ A.\ Hartnoll and R.\ Portugues,
{\it The chiral anomaly from M theory},
\hepth{0311088}.

\bibitem{Strassler03}
M.\ J.\ Strassler, {\it Non-supersymmetric theories with light scalar
fields and large hierarchies}, \hepth{0309122}.


\end{thebibliography}
\end{document}